\documentclass[11pt,twocolumn]{IEEEtran}

\usepackage{amsmath,mathrsfs,amssymb}
\usepackage[cmintegrals]{newtxmath}
\usepackage{bm} 
\usepackage{graphicx}
\usepackage{subcaption}
\usepackage{rotating}
\usepackage{soul}
\usepackage{color}

\begin{document}

\title{Extracting  Randomness From The Trend of IPI for Cryptographic Operators in Implantable Medical Devices}

\author{Hassan Chizari\footnote{Hassan is with University of Gloucestershire, UK}, Emil Lupu\footnote{Emil is with Imperial College london, UK}}

\maketitle

\begin{abstract}
Achieving secure communication between an Implantable Medical Device (IMD) inside the body and a gateway outside the body has showed its criticality with recent reports of hackings such as in St. Jude Medical's Implantable Cardiac Devices, Johnson and Johnson insulin pumps and vulnerabilities in brain Neuro-implants. The use of asymmetric cryptography in particular is not a practical solution for IMDs due to the scarce computational and power resources, symmetric key cryptography is preferred. One of the factors in security of a symmetric cryptographic system is to use a strong key for encryption. A solution to develop such a strong key without using extensive resources in an IMD, is to extract it from the body physiological signals. In order to have a strong enough key, the physiological signal must be a strong source of  randomness and InterPulse Interval (IPI) has been advised to be such that. A strong randomness source should have five conditions: \textit{Universality} (available on all people), \textit{Liveness} (available at any-time), \textit{Robustness} (strong random number), \textit{Permanence} (independent from its history) and \textit{Uniqueness} (independent from other sources). Nevertheless, for current proposed random extraction methods from IPI these conditions (mainly last three conditions) were not examined. In this study, firstly, we proposed a methodology to measure the last three conditions: Information secrecy measures for \textit{Robustness}, Santha-Vazirani Source $delta$ value for \textit{Permanence} and random sources dependency analysis for \textit{Uniqueness}. Then, using a huge dataset of IPI values (almost 900,000,000 IPIs), we showed that IPI does not have conditions of \textit{Robustness} and \textit{Permanence} as a randomness source. Thus, extraction of a strong uniform random number from IPI value, mathematically, is impossible. Thirdly, rather than using the value of IPI, we proposed the trend of IPI as a source for a new randomness extraction method named as Martingale Randomness Extraction from IPI (MRE-IPI). We evaluated MRE-IPI and showed that it satisfies the \textit{Robustness} condition completely and \textit{Permanence} to some level. Finally, we used NIST STS and Dieharder test suites and showed that MRE-IPI is able to outperform all recent randomness extraction methods from IPIs and its quality is half of the AES random number. MRE-IPI, still, is not a strong random number and could not be used as the secret key for a secure communication, however, it can be used as a one-time pad in exchanging the secret key for a communication. In this case, the usage of MRE-IPI will be kept at a minimum level and reduces the probability of breaking it. To the best of our knowledge, this is the first work in this area which uses such a comprehensive method and large dataset to examine the randomness of a physiological signal.
\end{abstract}

\section{Introduction}
Implantable Medical Devices (IMDs) provide a new perspective to healthcare system. Based on real-time date received from an IMD, the doctor could update the its settings using a gateway outside of the body of patient. This is a huge benefit for the patients as it helps the doctors for early detection of complications, early identification of the patients with high health risks and early recognition of suboptimal IMD functions \cite{Buchta:2017jx}. Moreover, with IMDs, the healthcare without affecting the patient's life becomes feasible (e.g. Diabetics and Dementia patients \cite{Walk:ie}). It also has been shown that remote monitoring of the patients using IMDs is very cost effective for the healthcare system as well \cite{GuedonMoreau:2014hd}. 

There are three classes of Implants: Fashion Trend, Life Enhancing and Life Preserving \cite{Walk:ie}. The first category belongs to the implants which designed to make the life more comfortable such as pet chips \cite{Anonymous:yv25OXTw} and key implants for unlocking doors or computers \cite{Hamill:7nAp32ic}. This category is usually based on RFID devices. Life enhancing implants are such as cochlear  or dental implants. This category of implants is mostly without battery, though a new generation of life enhancing implants are equipped with battery and sensors to collect information from the body and send it outside (e.g. collecting PH level and pressure in a knee \cite{XiaoyuLiu:2011ke}). The third category is life preserving implants. In this category, implants are mostly equipped with battery and in new generation of these devices, they are actively communicating with outside world and exchange information with their gateway device. The focus of this paper is on the life preserving IMDs which have sensing, computation and communication capability.    

Once an IMD is able to communicate by sending sensed data and receive operational commands, the security of the device becomes the main concern. The security of IMDs is one of the emerging research areas and many studies have highlighted its criticality in today's world. For instance, \cite{Pycroft:2016je} showed how an attack to a brain's IMD could take control of the implant and do some alteration on the victim's emotion. In another example, Reuters reported vulnerabilities in insulin pumps created by Johnson \& Johnson manufacturer \cite{Finkle:2016wy}, where attacker is able to intercept the commutation and/or request for a wrong dosage of insulin. More recently, Guardian \cite{Hern:2017uw} reported that more than half a million people are in danger because of the vulnerabilities found in St. Jude pacemakers. The challenge in secure communication between an IMD inside the body and a gateway outside the body is that the usage of asymmetric cryptography is not preferred \cite{Camara:2015bj,Kanjee:2014kk}. This is due to the resource hungry nature of public key cryptography, while IMDs are highly restricted in size \cite{AdamsJr:2016ek}, material \cite{Wu:2013br}, energy usage \cite{Chen:ir} and computation and communication power \cite{Cotton:2014ex}. Thus, many studies proposed symmetric encryption as a low resource hungry crypto-system for IMDs.

Based on the report published by WhiteScope \cite{Rios:2017wz}, more than 8000 vulnerabilities have been found in examining only seven models of peacemakers. The main reasons for this high number of vulnerabilities were, firstly, not using encryption method at all or secondly, using static (permanent) keys in encryption. To use temporary keys as a solution for this problem, IMD should be able to generate a strong cryptographic key each time the secure communication is needed. A great solution is proposed by \cite{Poon:2006ff} where in this method, IMD and the gateway measure a body physiological signal at the same time and create a secret key from it. Then, without any key exchanging process, they are sharing the same secret key. This method is called Physiological Value-Based (PVS) security solutions \cite{Venkatasubramanian:2010bw}. In this scheme, the security of the crypto-system highly depends on the randomness of physiological signal which is used to create a strong secret key. 

\begin{figure}
	\centering
	\includegraphics[width=\columnwidth]{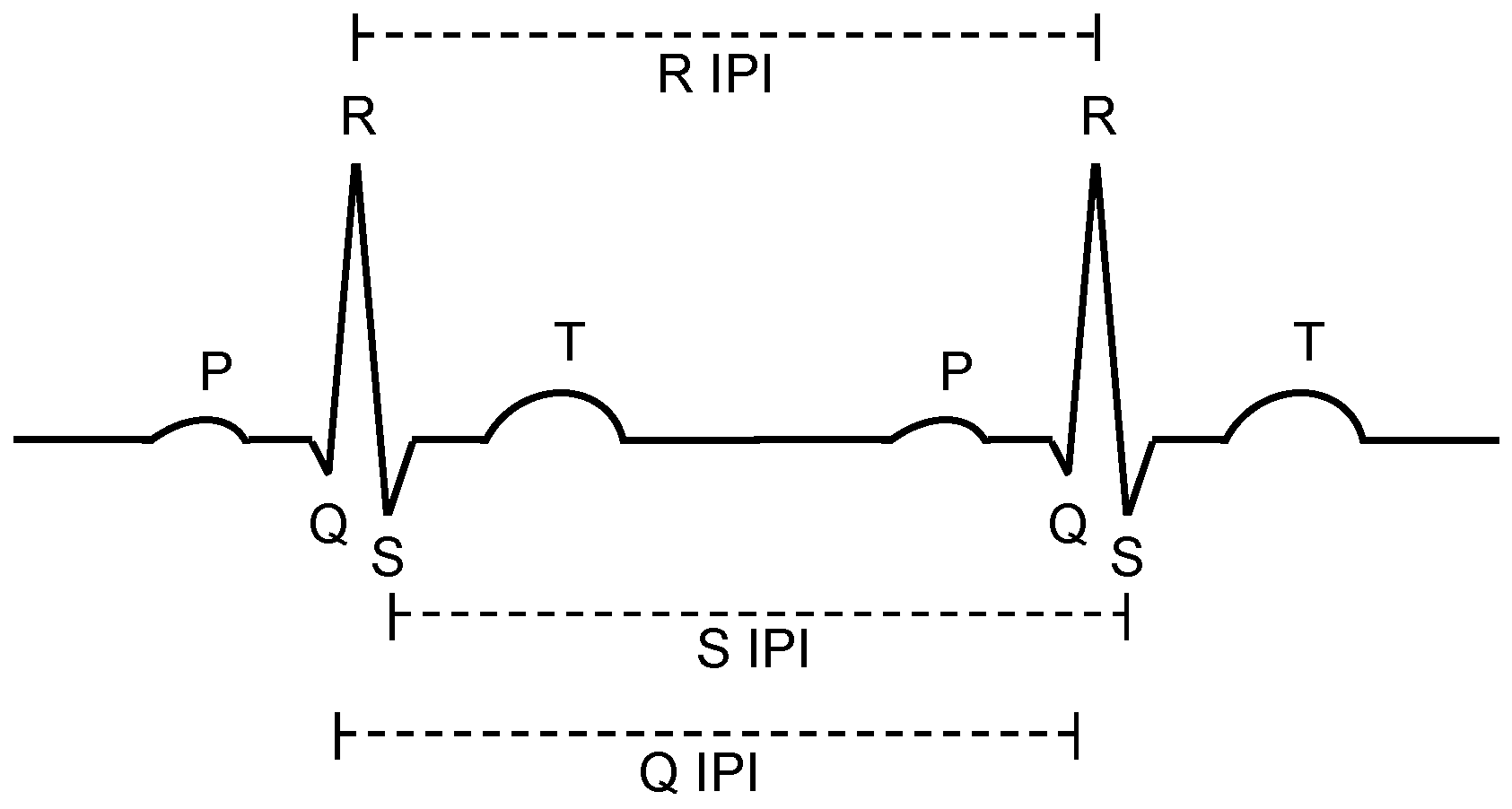}
	\caption{QRS peaks of Heartbeat (Source \cite{Peter:2016dv})}
	\label{fig:QRS}
\end{figure}

Several physiological signals have been proposed in the literatures to be used as the source of randomness for generating secret key such as Brain waves or electroencephalograms (EEG) \cite{Bajwa:2016ey}, electrocardiogram (EKG) \cite{Ali:2010gw} and Photoplethysmogram (PPG) \cite{Venkatasubramanian:2008ju}, Electrocardiography (ECG) \cite{Zheng:2014uh} and InterPulse Intervals (IPI) \cite{Poon:2006ff}. IPI is the time difference between two peaks of an ECG signal. As shown in Fig.~\ref{fig:QRS}, there are three peaks for every heart beat in the ECG signal named as Q, R and S. So, three IPI values could be extracted from the time difference between each two corresponding peaks (Q-Q, R-R, S-S). Among these, R has the highest peak and the easiest one to detect.  In the rest of this paper, whenever we refer to IPI, it is the R-R time difference. 

The contributions of this paper are, firstly, proposing a comprehensive methodology to examine the randomness of physiological signals. Although the focus of this paper is only on IPI, the methodology could be used on other signals as well. Secondly, using a large dataset with almost 900,000,000 IPI records, we showed that despite current claims, IPI is not a strong source of randomness. Thirdly, we developed a new randomness extraction method using Martingale Stochastic Process which is much stronger compared to other proposed methods in literature. To the best of our knowledge, this is the first work in this area which, introduces and uses a comprehensive evaluation method to examine the randomness strength of a physiological signal.   
 
This paper is organized as follows: in the next section, we discuss assumptions and requirements for the PVS method. Next, we present current randomness extraction methods from IPI. After that, we evaluate the randomness of IPI using a comprehensive methodology. Then, we present our proposed randomness extraction method followed by evaluating its strength compared to current methods. In final section, we conclude the paper by summing up the findings.   

\section{Requirements}
In PVS methodology, the gateway outside of the body initiates a request for communication to the IMD. Then, both devices start to measure a physiological signal. There are several assumptions here. Firstly, the physiological signal should be measurable from inside and outside the body. However, to preserve the privacy and security of IMD, the physiological signal should not be measurable with a distance to the body. This is based on the idea that if a gateway gets very close to the body, preferably touching the skin for a specific duration of time, it means that the gateway can be trusted. So, the physiological signal should be measurable from outside of the body but not from a distance. 

The second assumption in PVS methodology is that, measuring the PVS signal for implant does not consume considerable amount of energy. Since, the IMDs are very limited in every aspect of computing including energy, measuring physiological signal by them should be practically feasible. The third assumption is synchronization between the gateway and the IMD. In order to produce same secret key on both sides, IMD and gateway should start measuring the physiological signal at the same time. Based on the type of physiological signal a soft or hard synchronization is needed. Again, due to the limitations in IMDs, soft synchronization is more feasible as it needs less resources compared to a hard synchronization. After generating the secret key in both devices, if it is strong enough, it can be used as the session key. Otherwise, it can be used to setup a strong session key through the gateway. Among physiological signals, IPI can be easily measured inside and outside of the body. Touching the body would be enough for IPI and it does not need extra equipment for collecting signals (e.g. EEG signal). IPI can be collected with soft synchronization, since after handshaking, both devices need to wait for the first heartbeat R-peak to start the measurement. The question remains is that how strong will be the secret key generated from IPI.  

A physiological signal must be a good source of randomness for generating strong secret key by satisfying five conditions \cite{Ramli:2013wp}. The first condition is the \textit{Universality} referring to the availability of the signal on all people. The second condition is \textit{Liveness} where the physiological is always available for measurement any time anywhere. Thirdly, the extraction of a new secret key should be always possible from the physiological signal (\textit{Permanence}). For instance, if the secret key $s_1$ is generated from a physiological signal at the time $t_1$, $s_n$ generated at time $t_n$ should not be guessable based on $\{s_1,s_2,...,s_{n-1}\}$. If a physiological signal does not have the feature of \textit{Permanence}, it can only be used once or very occasionally, because repetitive usage of it provides enough information for an adversary to guess it. The next condition is \textit{Uniqueness}. It means that the physiological signal on two different subjects should be accounted as two independent sources of randomness. Considering the availability of physiological signals for the subject $i$ ($p_i=\{s_{i,1},s_{i,2},...,s_{i,n}\}$), $s_{j,n}$ which is physiological signal at time $n$ for subject $j$, should not be guessable based on $p_i$. \textit{Robustness} is the last condition which shows that computationally guessing the secret key should not be feasible whether or not the adversary has some information about the physiological signal. 

\textit{Robustness} of a source is calculated by information secrecy measures. Information secrecy measures examine the strength of the secret key against an adversary who tries to guess the key. Based on the knowledge of the adversary from the secret key, three scenarios could happen. In the best-case scenario, adversary does not have any information about the secret key and wants to blindly guess it. This situation is called \textit{Perfect Secrecy}. In the second scenario which is called \textit{Conditional Secrecy}, adversary has some information correlated with the key. \textit{Unconditional Secrecy} is the worst-case scenario where adversary knows the distribution histogram of the secret keys. This helps adversary to find which values could be the best guess for determining the secret key. For each scenario, there is a method to measure the strength of the secret key. In addition to these three, \textit{Probabilistic Bound} is the forth information secrecy measure which shows the distance between the secret key distribution and a uniform distribution \cite{Wolf:2003bf}.   

Among these physiological signals, IPI has some unique properties which makes it one of the most cited methods in the literature to be used as the source of randomness in generating the secret key. Firstly, it has the condition of \textit{Universality} as everyone has the heartbeat signal. There are two exceptions here. First one is the flat line heart pulse in emergency situation and the second one is the controlled heartbeat by a peacemaker. Both of these situations are excluded from the scope of this article. The second property of IPI which makes it a good source of randomness for generating the secret key is that, it has the \textit{Liveness} condition as well, where the heartbeat signal is always available to measure for everyone (excluding two situations which mentioned above). Thirdly, detection of IPI does not need complex hardware or software analysis and is very simple to be implemented in an IMD. Lastly, it can be measured in almost every location inside the body and outside by touching the skin. The only question remains is if the IPI is a strong source of randomness for generating secret key to be used in a crypto-system. Although it has been claimed in many studies that IPI is a strong randomness source, its \textit{Permanence}, \textit{Uniqueness} and \textit{Robustness} have not been examined. 

\section{Related Work}
Using IPI as a source of randomness has been proposed by \cite{Poon:2006ff}, where a random extraction algorithm is needed to convert IPI value to a random number. A few randomness extractor algorithms from IPI have been proposed in the literature. Proposed methods are including using XOR function \cite{Wang:2010jg}, gray-coding \cite{TianHong:2011ez} and using frequency domain \cite{Ramli:2013dq, Ramli:2013wp, Kalaivani:2015jc}. In some studies, a combination of algorithms is used for randomness extraction. For instance, \cite{Bao:2012fw,Miao:2012cb,Bao:2013jd} used accumulation, modulo, contract mapping and gray-coding for the extractor and \cite{Altop:2015ej} proposed a combination of concatenation, quantization and gray-coding as the randomness extractor from IPI. 

Whatever the extraction method is, it needs to be evaluated for the quality of generated randomness. Several extracting methods did not perform any randomness test to examine the quality of proposed algorithm (e.g. \cite{Cho:2012fp, Hu:2013hs, Ramli:2013dq, Ramli:2013wp, Jammali:2015ii, Zheng:2015bx, Kalaivani:2015jc}). In these works the aim of the paper was not examining the randomness of proposed extraction method. There are other randomness extraction works in which there was an attempt to measure the quality of proposed method. However, in these works, two aspects have been somewhat ignored. To evaluate the quality of a randomness extractor, the dataset and the methodology of evaluation are the key points. For instance, \cite{TianHong:2011ez} evaluated the randomness property of their algorithm with 5 minutes ECG data of 10 subject by NIST Statistical Test Suite (STS) \cite{Rukhin:2000ee} randomness test. \cite{Bao:2012fw,Miao:2012cb,Bao:2013jd} tested their algorithm using 5 minutes ECG signal of 40 subjects with NIST STS. In another work, \cite{Zheng:2014uh} used histogram analysis on 1500 consecutive IPI values. \cite{Seepers:2014ea} tested their proposed method with 100 subjects' ECG data with Entropy test. \cite{Altop:2015ej}, to evaluate the randomness of proposed algorithm, used Temporal Ratio \cite{Israel:2005fh} method over 5 minutes ECG data of 50 subjects. 

Current evaluation methods for examining the randomness of IPI are limited to using a few randomness tests functions against a very small dataset of IPI values. In order to have a complete evaluation of the strength of a randomness source, first of all, conditions such as \textit{Permanence}, \textit{Uniqueness} and \textit{Robustness} (including \textit{Perfect Secrecy}, \textit{Conditional Secrecy}, \textit{Unconditional Secrecy} and \textit{Probabilistic Bound}) of the random source must be examined. Secondly, a large data set of IPIs is needed for evaluation of aforementioned measures. In all previous works, the maximum number of subjects to examined were 100 and the largest number of IPI reading from one subject was 1500 consecutive values. This is far below the number of IPI values which are needed to examine the strength of randomness. For instance, to measure the \textit{Unconditional Secrecy} in \textit{Robustness} condition (Min-Entropy of a string with size 16 which will be discussed later), considering the best-case scenario where the distribution of the source is uniform, at least $1000*2^{16}=65,536,000$ samples of IPIs are needed to provide a confidence interval of $95\%$ in the result. None of the current randomness extraction methods from IPI used such methods and large datasets to evaluate their algorithms. In contrast to previous works, we gathered a large dataset of IPI values. Moreover, we proposed methods for measuring all the conditions to examine the strength of the IPI as a randomness source. In addition to that, we developed a randomness extraction algorithm and compared its strength against these conditions and reported it.  


\section{IPI Randomness}
\subsection{Dataset}
In this work, using PhysioNet \cite{Goldberger:2000br}, we have created a dataset of IPIs containing 4338 subjects and a total of 895,621,566 IPI values. This dataset comprises of a wide range of subjects including healthy subjects resting, healthy subject during Cardio test, subjects with heart problem and failure, infants, children, young, middle age and senior subjects both male and female. We did not restrict the dataset selection from PhysioNet's datasets and the aim was to collect as much as possible IPI data. Whenever possible, the IPI signal is get from audited signal files (ATR files), or from non-audited annotation files (QRS, WQRS, ECG and XYZ files), otherwise using QRS procedure of PhysioNet in Matlab, IPIs are extracted from DAT files.

\begin{figure}
\centering
\includegraphics[width=\columnwidth]{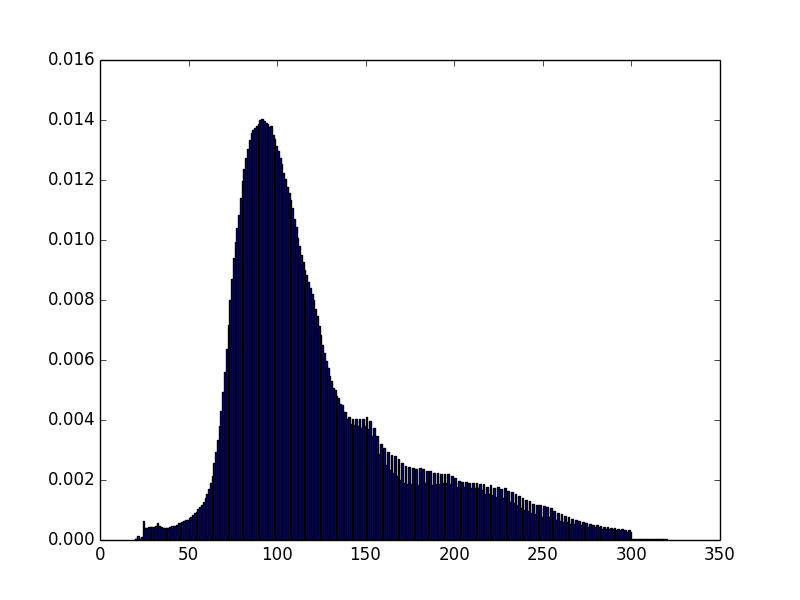}
\caption{Histogram analysis of IPIs}
\label{fig:hist}
\end{figure}

We have extracted the IPI distribution histogram from the harvested dataset is presented in Fig.~\ref{fig:hist}. The minimum value of IPI in this histogram is 20 which is equivalent to $300bpm=60/0.20$ (heartbeat per minute) and the maximum is 330, equivalent to $18bpm=60/3.30$. To convert an IPI value to a binary representation, an 8-bit string is used for conversion of $ipi = IPI-20$. We subtracted IPI from 20, so $ipi$ starts from 0. Thus, the $ipi$ value will be from 0 to 310 which to present it in binary, 9 bits are needed. However, as it is shown later, the most significant bit in $ipi$ does not provide much randomness and has been removed. So, we used an 8-bit representation of $ipi$ value in following sections. 

\subsection{String Size}
To apply the measures of \textit{Permanency}, \textit{Uniqueness} and \textit{Robustness}, we need to consider the string size of concatenated measures. Considering a series of IPI values, we define the string of $Z_k=[ipi_1 \cdot ipi_2 \cdot ... \cdot ipi_n]$ as the concatenation of $ipi_j=\{0,1\}^k$ where $k$ is the number of bits from each $ipi$ value ($0<k \leq 8$). For example, consider a series of IPI values as 160, 125, 132, 171, 148 and 130, then deducting 20 will give us $ipi=\{140,105,112,151,128,110\}$ and using only the two least significant bits of each $ipi$ ($k=2$) and concatenating them, we have $Z_2=[000100110010]$. In order to examine the randomness of $Z_2$, all the measurement methods need the distribution of bits in $Z_2$. If we look at $s=\{0,1\}^1$ for $Z_2$, we have $D_{{Z_2},1}=\{8,4\}$, which means 8 zeros and 4 ones. For distribution of the string $s=\{0,1\}^2=\{00,01,10,11\}$, there are four combinations to look at. To calculate $D_{{Z_2},1}$, $Z_2$ is parsed twice. In the first parse, we have $a_{{Z_2},1}=\{00,01,00,11,00,10\}$. In order to make sure that all possible patterns of bits in $Z_2$ have been observed, we use circulation method which has been used in several randomness test suites (e.g. \cite{LEcuyer:2007ki,Brown:2013wb}) for creating the distribution histogram. Thus, we go to the second parse and rather than starting from the first bit in $Z_2$, we start from the second bit and then we have $a_{{Z_2},2}=\{00,10,01,10,01,00\}$. For the last bit in $Z_2$, we concatenate it with the first bit of $Z_2$ to have $00$ at the end of $a_{{Z_2},2}$. Now, looking at both $a_{{Z_2},1}$ and $a_{{Z_2},2}$, we have $D_{{Z_2},2}=\{5,3,3,1\}$. With the same process, when $s=\{0,1\}^3$, three parses on $Z_2$ will be done and the distribution of $Z_2$ will be $D_{{Z_2},3}=\{2,3,2,1,3,0,1,0\}$. In theory, to calculate the distribution for string $s=\{0,1\}^n$, $n$ must be $1 \leq n < \infty$, however, in practice the maximum size of $n$ is associated with the size of dataset. As discussed in the first section, the maximum value for $n$ in this article is $16$.

\subsection{Robustness}
IPI as a source of randomness has the robustness property if knowing  its probabilistic distribution (partially or fully) would not help in predicting its next value. Based on the information secrecy measures \cite{Wolf:2003bf},  we measure \textit{Perfect Secrecy} by Shannon Entropy (Eq. \ref{eq:shannon}) \cite{Shannon:2013iy} which quantifies the encoded length of the source. In this mode, adversary has no knowledge from the distribution of the source. We measure \textit{Conditional Secrecy} using Renyi Entropy or its descendant Collision Entropy (Eq. \ref{eq:collision}) \cite{Cachin:1997wn} which bounds the collision probability between samples. The second measure of \textit{Conditional Secrecy} is Guessing Entropy (Eq. \ref{eq:guess}) \cite{Massey:gs} which shows the difficulty of guessing the value of a random variable. \textit{Unconditional Secrecy} quantifies unpredictability of the source and we measure it using Min-Entropy (Eq. \ref{eq:min})\cite{Espinoza:2013cd}. Indeed, in \textit{Unconditional Secrecy}, adversary has complete knowledge of the distribution histogram of the random source. So, it knows which output from the random source has the highest probability to occur. The last information secrecy measure is \textit{Probabilistic Bounds} which calculates the distance between distributions of the random source and uniform distribution over the same range (Eq. \ref{eq:bound}). Considering $X \in \{0,1\}^n$, a randomness extractor works as Ext : $\{0,1\}^n \rightarrow \{0,1\}^k$ such that $Ext(X)$ is distributed in $\{0,1\}^k$, then the entropies of a random variable $X$ is defined as:

\begin{eqnarray}
\label{eq:shannon}H_{Sh}(X) = -\sum_{x \in R(X)} Pr[X=x] . log(Pr[X=x]) \\
\label{eq:collision}H_{\alpha=2}(X) \equiv \frac{1}{1-\alpha} \sum_{x \in R(X)} Pr[X=x]^{\alpha} \\
\label{eq:guess}G(X) = \sum_{x \in R(X)}Pr[X=x](x+1) \\
\label{eq:min}H_\infty(X) = \underset{x \in R(X)}{min}\bigg\{log\frac{1}{Pr[X=x]}\bigg\} \\
\label{eq:bound}||P_x - P_y||_1 = \sum_{x \in R(X)}|P_x[X=x] - P_y[X=x]|
\end{eqnarray}
 
Many researchers used all the 8 bits of an IPI (e.g. \cite{Cho:2012fp}) as the source of randomness, however, there are some works only considering a subset of it. For instance, \cite{Hu:2013hs,Jammali:2015ii} used the first 4 bits of an IPI and \cite{Seepers:2014ea} used the first 2 bits of the IPI. To identify which combination of bits of an IPI is the best source for randomness, in all measurements, we examined 8 scenarios which in the first scenario only one bit (the least significant bit) is the source of randomness. Using respectively for all scenarios, in scenario $n$ ($n \in \{1..8\}$), the first $n$ least significant bits of IPI is used as the source of randomness. Corresponding to these 8 scenarios, we created 8 datasets from the main $ipi$ database. The first dataset contains the first least significant bit of an IPI. Dataset No. 2 contains the two least significant bits of an IPI and so on and so forth till Dataset No. 8 which contains all the 8 bits of an IPI value.
 
\begin{figure}
\centering
\includegraphics[width=\columnwidth]{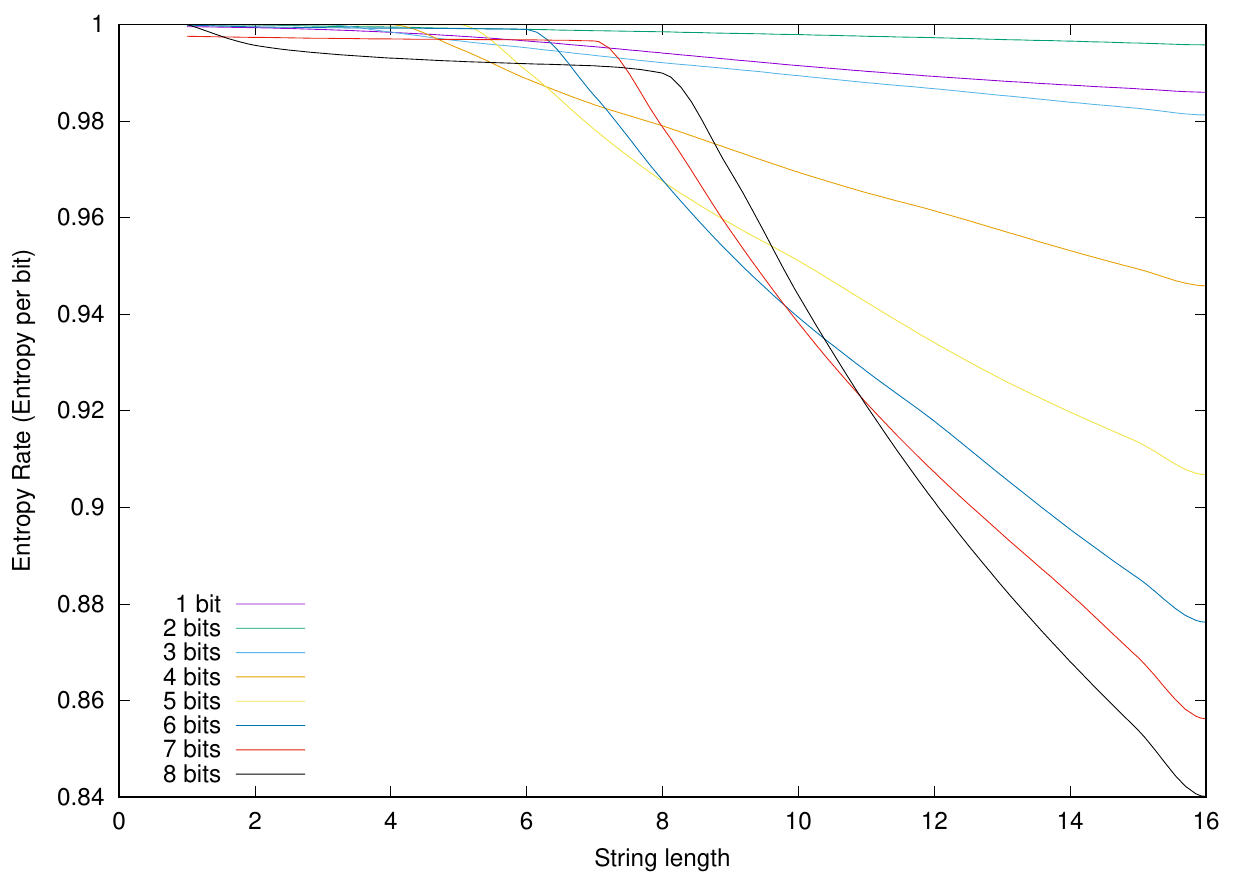}
\caption{Shannon Entropy of selected bits of IPI in various string lengths}
\label{fig:entropyshannon}
\end{figure}

The result of Shannon Entropy analysis of IPI is presented in Fig.~\ref{fig:entropyshannon}. As shown, the dataset which contains the two least significant bits of IPI (Dataset 2) has the highest Shannon Entropy value. It is also clear that even by increasing the length of string in Entropy calculation, the Shannon Entropy of dataset 2 is still close to one. The worst Entropy belongs to dataset 8 where all 8 bits of IPI is being used. 

Since dataset 2 has the Shannon Entropy of almost 1, in the condition of \textit{Perfect Secrecy} where adversary has zero knowledge about the random source distribution, it has almost perfect random quality. The next step is to measure the entropy when the adversary has partial information about the random source (\textit{Conditional Secrecy}).

\begin{figure}
\centering
\includegraphics[width=\columnwidth]{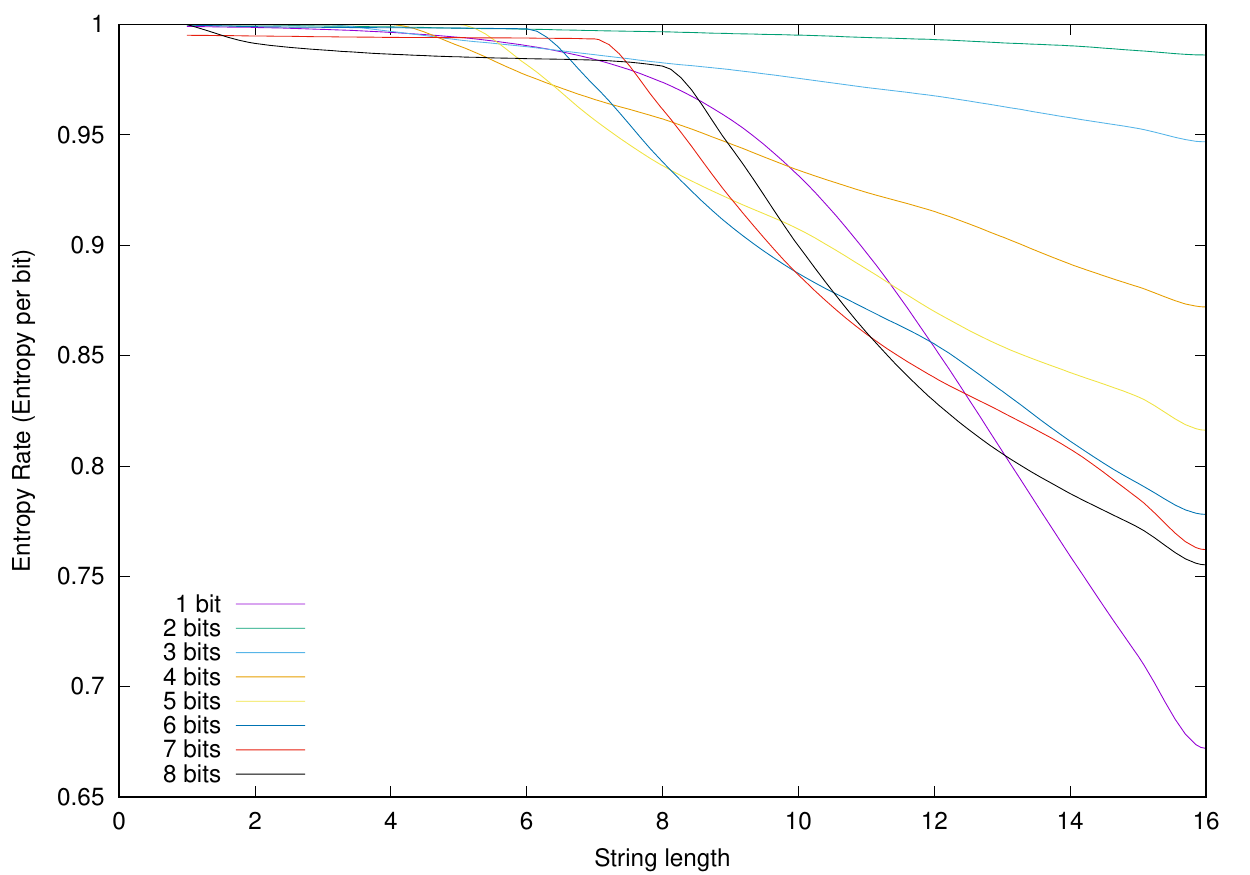}
\caption{Collision Entropy of selected bits of IPI in various string lengths}
\label{fig:entropycollision}
\end{figure}

Fig.~\ref{fig:entropycollision} shows Collision Entropy results which is almost similar to Shannon Entropy with a small decrease. Once again, dataset 2 has the highest entropy value and up to the string length of 16 is still over 0.95. Collision Entropy shows the diversity of string patterns in the dataset and for this, dataset 2 shows a very high diversity. Collision Entropy or in general Renyi Entropy is a measure for \textit{Conditional Secrecy} where adversary knows a value which has some correlation with the random value. In this situation, also, dataset 2 shows an almost perfect form of randomness.

\begin{figure}
\centering
\includegraphics[width=\columnwidth]{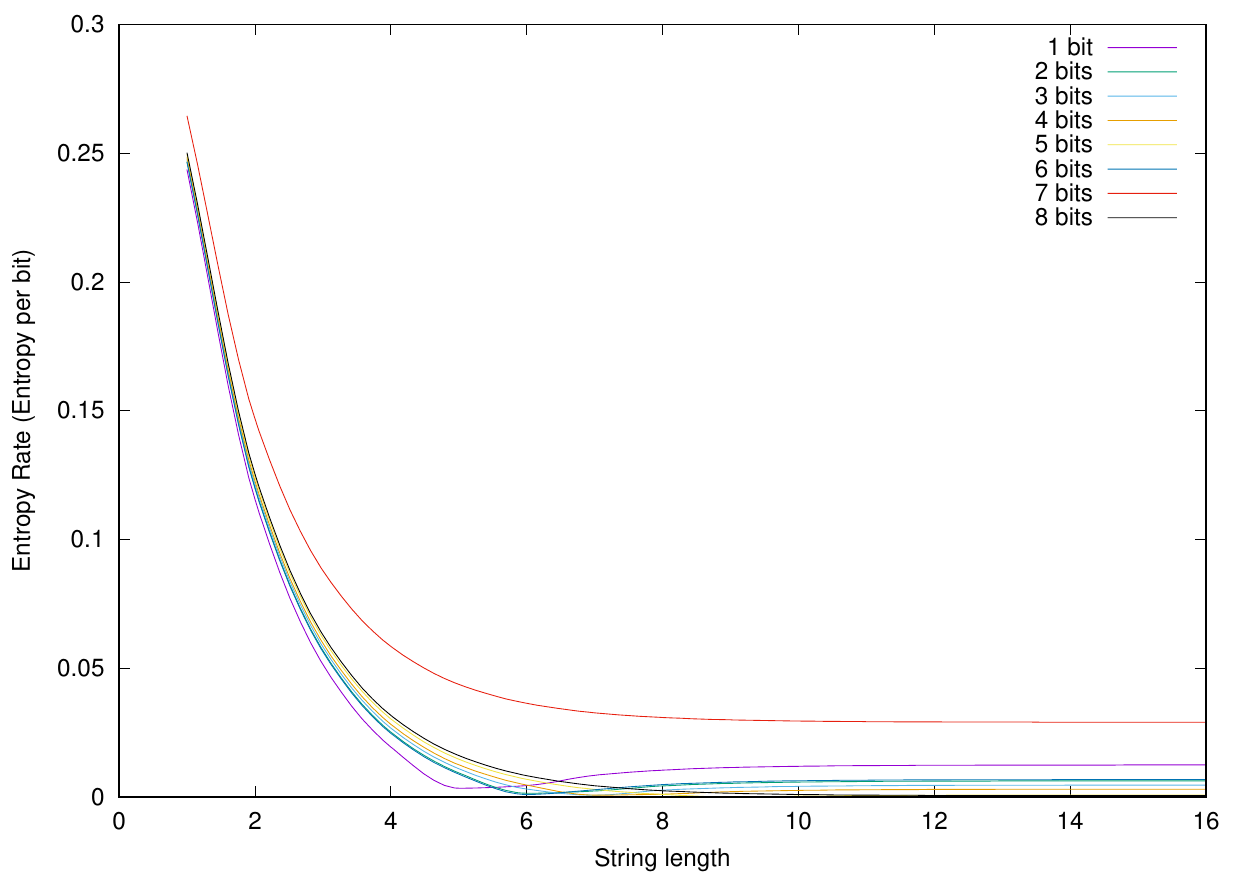}
\caption{Guessing Entropy of selected bits of IPI in various string lengths}
\label{fig:entropyguess}
\end{figure}

The second measure of \textit{Conditional Secrecy} is Guessing Entropy (Fig.~\ref{fig:entropyguess}), the value of entropy is subtracted from 0.5. The reason is that guessing entropy shows the difficulty of guessing the random value and the entropy rate of guessing entropy is the probability of guessing the right value for a bit. So, the value 0.5 for guessing entropy rate shows that the adversary is not able to have a good guess about the bit and the chance to be correct is 50\%. Thus, we used the 0.5 as the benchmark line and deduct the guessing entropy rate from it and now a value of zero in the curve means that the probability of guessing the next bit is 0.5. By increasing this value, the probability of guessing the next bit is increasing. As shown, for all datasets, Guessing Entropy (subtracted from 0.5) is very close to zero which means the probability of guessing the next output from the source is 0.5. In this test, closest values to 0 are for datasets 2,3,4 and 5.

\begin{figure}
\centering
\includegraphics[width=\columnwidth]{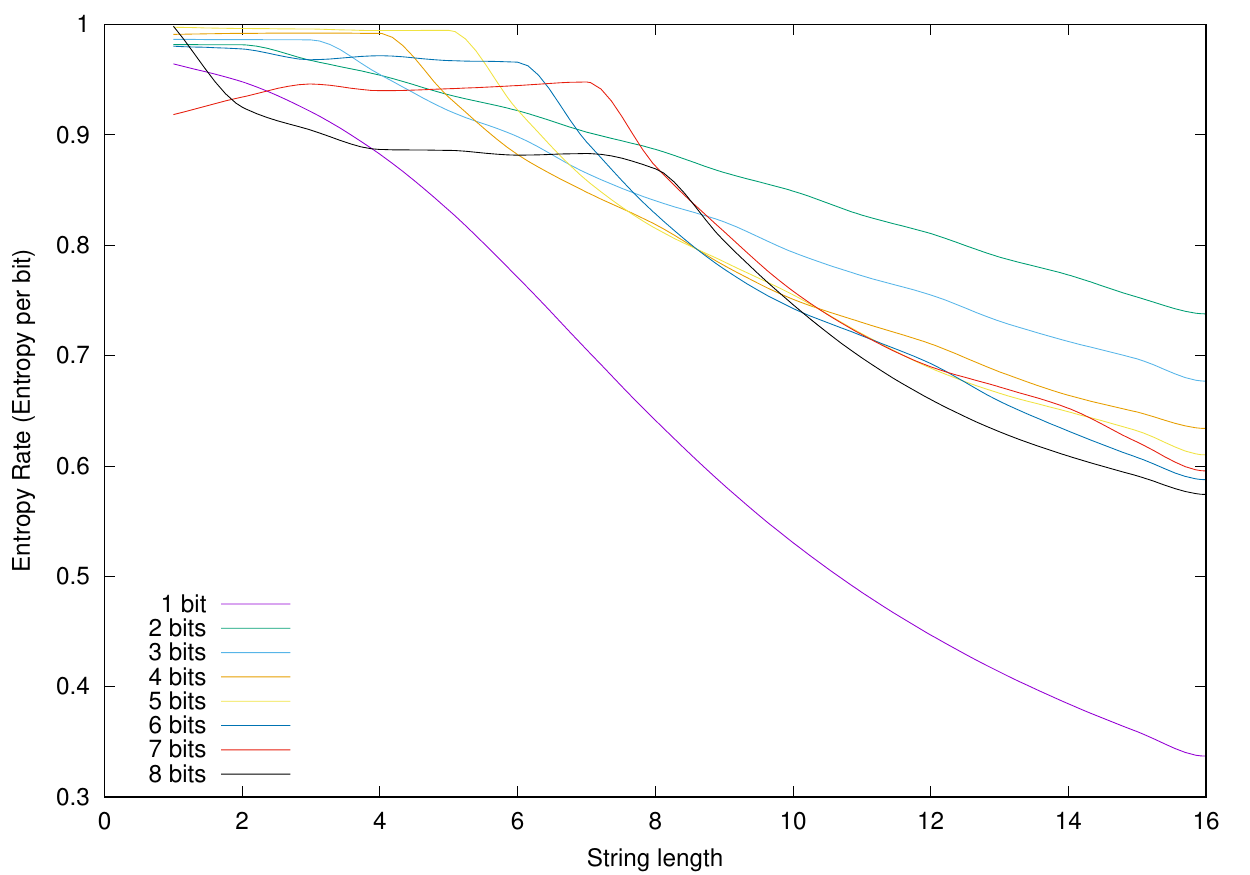}
\caption{Min-Entropy of selected bits of IPI in various string lengths}
\label{fig:entropymin}
\end{figure}

Min-Entropy is the measure of \textit{Unconditional Secrecy}, where the adversary knows the distribution histogram of the random source. Min-Entropy result (Fig.~\ref{fig:entropymin}) shows that IPI is not performing well in almost all datasets. As shown, the value of entropy rate drops to lower than 0.8 by increasing the string size to 16 in all datasets. However, dataset 2 still has the highest Entropy value compared to others. Min-Entropy is a measure of predictiveness of the string where values close to one indicate unpredictability of the source and for IPI it demonstrates that the source is not highly unpredictable. 

\begin{figure}
\centering
\includegraphics[width=\columnwidth]{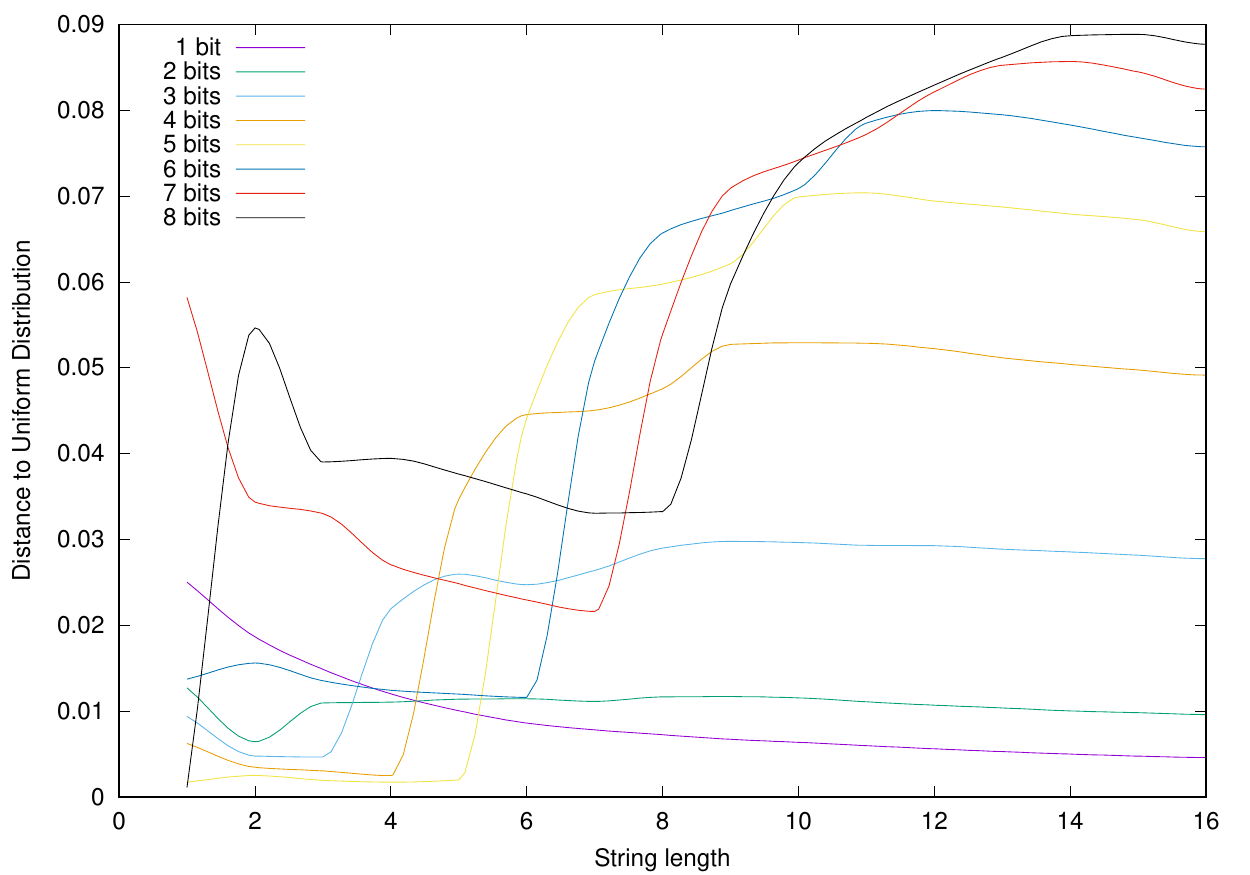}
\caption{Uniformity Check for selected bits of IPI in various string lengths}
\label{fig:pbound}
\end{figure}

Last step in analysing the \textit{Robustness} of randomness source is measuring the \textit{Probabilistic Bound}. In this process, we calculate the distance between random variable and a uniform distribution using Eq.~\ref{eq:bound}. The closer value to zero shows smaller distance to a uniform distribution which means better value for \textit{Probabilistic Bound}. As shown in Fig.~\ref{fig:pbound}, the closest dataset of IPIs to uniform distribution belongs to dataset 1, 2. Especially, in dataset 1, in all string lengths the cumulative distance to a uniform distribution is less than $0.1$. 

The \textit{Robustness} analysis of IPI using the information secrecy measures reveals two points. Firstly, if the adversary has no knowledge or limited knowledge about the distribution of IPI, IPI can be used as a very robust source of randomness. However, it seems reasonable to consider that the histogram of IPI distribution is not hard to collect (see Fig.~\ref{fig:hist}). If the adversary has the knowledge about the IPI distribution histogram, then IPI is not a very robust source of randomness despite the claim made by many researchers lately. The second point from \textit{Robustness} analysis is that the dataset 2 which contains the two least significant bits of IPI values, is a more robust randomness source compared to other datasets. 

\subsection{Permanence}
IPI as a randomness source has \textit{Permanence} condition if knowing the history of the heartbeat of one person would not help in predicting his/her future heartbeats. To measure the \textit{Permanence}, we propose using Santha-Vazirani method. A randomness source is called Santha-Vazirani source (SV-source) \cite{Santha:1986gd} where the outcome of last generated bit is related to the previous outcomes. For source $X$ and $\delta \in [0,1]$, we have:

\begin{equation}
\forall i \in n, \forall x_i \in \{0,1\} \rightarrow \frac{1-\delta}{2} \leq Pr[X_i = x_i | \forall x_{i-1}] \leq \frac{1+\delta}{2}
\end{equation}
$\delta$ is the bias for the new bit $x_i$, which it has some dependencies to the previous bits in the source $\{0,1\}^{i-1}$. In a simpler form we have:
\begin{equation}
\forall x,y \in \{0,1\}^n \rightarrow \frac{Pr[X=x]}{Pr[Y=y]} \leq \frac{1+\delta}{1-\delta}
\label{eq:santha}
\end{equation}

\begin{figure}
	\centering
	\includegraphics[width=\columnwidth]{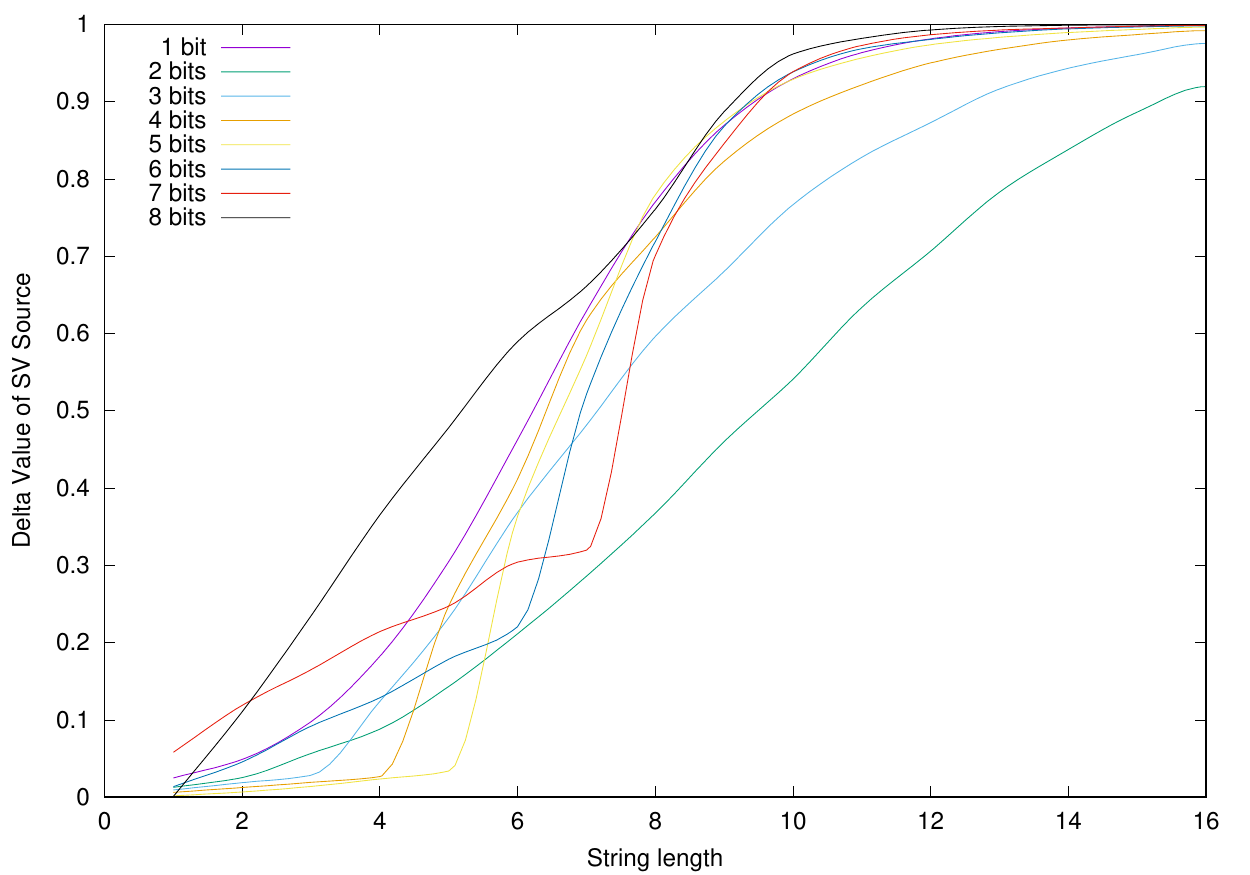}
	\caption{$\delta$ value of SV-source for various string lengths of IPIs}
	\label{fig:delta}
\end{figure}

The best possible $\delta$ value is zero for any string length which demonstrates that the source is not Santha-Vazirani. If $\delta$ is equal to zero, the probability of having zero or one is always 0.5, no matter how much data is available. To evaluate the predictiveness of source $X$ from Eq.~\ref{eq:santha}, we calculated the maximum and minimum of $Pr[X=x]$ for $\forall x \in \{0,1\}^n$ where $n = 1..16$ using the 8 datasets of IPI values. Then, using Eq.~\ref{eq:santha}, we calculated the $\delta$ value of SV-source and results are presented in Fig.~\ref{fig:delta}.

As shown in Fig.~\ref{fig:delta}, IPI is a SV-source as there is a great dependency for the new outcome of the source to its history. Once again, dataset 2 has the lowest $\delta$ value compared to other datasets. However, for all the datasets, when the string length is more than 10, $\delta$ is almost equal to one. That means that with enough history, the adversary is able to predict the next random value with the probability very close to 1.  

It has been mathematically proven that the extraction of a uniform random bit from a SV-source is impossible \cite{Santha:1986gd}. By this, we conclude that the IPI value has not the condition of \textit{Permanence} as a source of randomness to be used in a crypto-system. 

\subsection{Uniqueness}
IPI as a randomness source has the \textit{Uniqueness} condition if having the heartbeat of someone would not help in predicting the heartbeat of another person. For short, the random sources should be independent from each other. To measure the dependency of different subjects IPI values from each other, we propose to use the dependency analysis of random sources presented by \cite{Konheim:1981:CP:539492}. As shown in Eq.~\ref{eq:dependency1}, the summation of entropies of two randomness sources are bigger or equal to the entropy value of their combination as one source. In Eq.~\ref{eq:dependency2} two sources are independent from each other only and if only $lim e_{indp} \rightarrow \infty$.   

\begin{eqnarray}
H(X,Y) \leq H(X) + Y(Y) \\ \nonumber
H(X) = -\sum_t p_x(t)\log p_x(t) \\ \nonumber
H(Y) = -\sum_t p_y(s)\log p_y(s) \\ \nonumber
H(X,Y) = \sum_{t,s}p_{x,y}(t,s) \log p_{x,y}(t,s) \label{eq:dependency1}
\end{eqnarray}
 
\begin{equation}
e_{indp} = H(X) + H(Y) - H(X,Y)
\label{eq:dependency2}
\end{equation}

\begin{figure}
	\centering
	\includegraphics[width=\columnwidth]{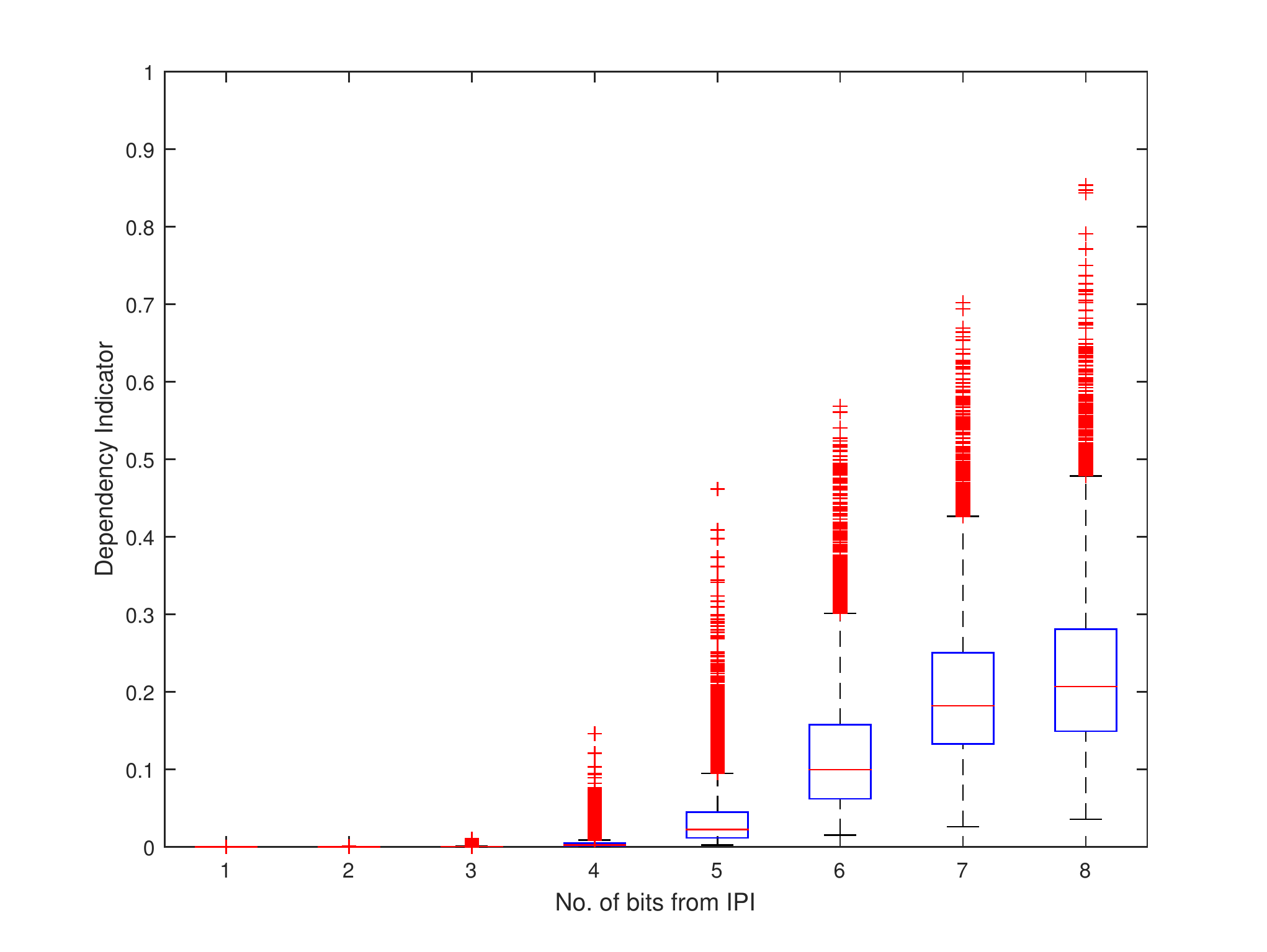}
	\caption{Dependency analysis among IPIs from different subjects}
	\label{fig:dependency}
\end{figure}

To examine the independence of IPI values of differentness sources to each other, we selected 1360 subjects from the IPI dataset. These subjects have more than 100,000 consecutive IPI values because dependency analysis between random sources needs a large sample size. From this pool of 1360 subjects, for 10,000 times, we selected two random subject and calculate the $e_{indp}$. The box plot of $e_{indp}$ values is presented in Fig.~\ref{fig:dependency}. As shown, datasets of 1, 2 and 3 are showing the condition of \textit{Uniqueness} completely. For dataset 4, the average value of $e_{indp}$ is almost zero, despite a few outliers. For other datasets, the dependency between two random sources is shown with high average values and outliers.  

In conclusion, despite the claim has been made by previous works, the value of IPI is not a good source of randomness. It provides \textit{Robustness} to some level, but considering the worth case scenario (\textit{Unconditional Secrecy}), it fails to provide \textit{Robustness}. Moreover, IPI could not satisfy the \textit{Permanence} condition as it is highly dependent to its history. Finally, IPI provides the \textit{Uniqueness} as a random source which shows every heart could be considered as an independent source compared to others. Among the 8 datasets used in this experiment, dataset 2 has better results, which shows the two least significant bits of IPI are the better source of randomness compared to other combination of bits. However, due to the fact discussed, even the two least significant bits of IPI is not a strong random number and thus not appropriate to be used in a crypto-system.

\section{Martingale Randomness Extraction for IPI (MRE-IPI)}
Rather than using the IPI values, which as we have shown above are not a good source of randomness, we propose to use the trend in which IPI is changing. Intuitively, whilst the value of IPI is affected by its previous values, the trend and rate of change are influenced by both physiological and contextual factors, activities, emotions and etc. Martingale Stochastic Process \cite{Downey:2004jxa} has been used in many applications of randomness such as extraction \cite{Beigi:2014ue}, developing computable randomness \cite{Rute:2016br}, algorithmic randomness theory \cite{KjosHanssen:2014ht} or even stock market analysis \cite{JohnAyodele:2017ct}. However, it has not been used to extract the randomness from IPI trend. 
 
Martingale is a sequence of random variable where the expectation of the next value is equal to the previous value. There are two main properties for Martingale as follows:

\begin{eqnarray}
\forall n, \mathop{\mathbb{E}}(|X_n|) < \infty \\
\forall n, \mathop{\mathbb{E}}(X_{n+1}) = X_n
\label{eq:martingaleproperty}
\end{eqnarray}

As shown in Eq~\ref{eq:martingaleproperty}, in Martingale Stochastic Process the expectation of next value is equal to the previous value. In order to use Martingale on IPI, we must convert it to the Martingale Stochastic Process. Then, we use this Martingale and extract the random variable from it. We called this algorithm as Martingale Randomness Extraction for IPI (MRE-IPI). To develop stochastic process, extract the randomness and test it, we used dataset 2 from previous section which contains the least two significant bits of IPI. From now on, for simplicity we refer to dataset 2 as only \textit{the dataset}. Moreover, to answer that MRE-IPI is not biased on the IPI values in the dataset, the dataset is divided to two subsets randomly, one for randomness extraction analysis (training dataset) and another one for evaluating the proposed method (testing dataset). The train dataset consists of 2212 subjects with the total of 457,491,012 IPI values and the test dataset consists of 2126 subjects with the total of 438,130,554 IPI values. 

The first step in MRE-IPI is to develop a conversion function $\mathscr{F}:\{0,1\}^t \rightarrow \{0,1\}$ which can satisfy Eq.~\ref{eq:martingaleproperty}. In this situation, we prefer to use the minimum possible value for $t$. Using of large values for $t$, increases the number of IPI values needed to generate one random bit. For example, if $t=20$, using 2 bits of an IPI, we need 10 IPI values to generate one bit in Martingale Stochastic Process. Based on our analysis the optimum value for $t$ is equal to 3. To do this, we calculated the probability distribution for all possible combinations of 3 bits for train dataset as presented in Table~\ref{tbl:strings}.

\begin{table}
\centering
\small
\caption{Probabilistic distribution of all possible combination of bits for string length of 3}
\begin{tabular}{| l |l | l |}
\hline
No. & String & Probability of occurrence \\ \hline
0 & 000 & 0.133567088 \\ \hline
1 & 001 & 0.122682815 \\ \hline
2 & 010 & 0.130448785 \\ \hline
3 & 011 & 0.119522821 \\ \hline
4 & 100 & 0.122658224 \\ \hline
5 & 101 & 0.127181921 \\ \hline
6 & 110 & 0.119505945 \\ \hline
7 & 111 & 0.1244324 \\ \hline
\end{tabular}
\label{tbl:strings}
\end{table}

The best combination of strings which could provide the commutative probability of occurrence of 0.5 is as follow:

\begin{itemize}
	\item Group 1: $G_1 = \{000,011,101,110\}$, where $P(X=G_1) = 0.499777776$
	\item Group 2: $G_2 = \{001,010,100,111\}$, where $P(X=G_2) = 0.500222224$
\end{itemize}

Consider $X_n=x$ as a random variable and $X_{n+1}$ as the next random variable in a time series, then $X \leftarrow \mathscr{F}(s)$ is defined as:

\begin{equation}
X_{n+1} = \begin{cases}
X_n + 1 & s \in G_1 \\
X_n - 1 & s \in G_2 \\
\end{cases}
\label{eq:martingaleS}
\end{equation}
where $s \in S=\{0,1\}^3$, and $S$ is the random variable of IPI from training dataset 2. Then $E\mathop{\mathbb{E}}(X_{n+1}) = G_1 * (x+1) + G_2 * (x-1)$. As $G_1 , G_2 \approx 0.5$, then $\mathop{\mathbb{E}}(X_{n+1}) = x$, which is the value for $S_n$. Thus, $\mathop{\mathbb{E}}(X_{n+1}) = X_n$ and Eq.~\ref{eq:martingaleS} is a Martingale Stochastic Process (Eq.~\ref{eq:martingaleproperty}).

The second phase is to use $X \leftarrow \mathscr{F}(s)$ with two threshold values to produce the random bit. Consider $t_1 > 0$ and $t_2 < 0$ as two threshold points, then we have:

\begin{eqnarray}
	\forall n, X_n \leftarrow \mathscr{F}(s) \\
	X_n > t_1 \rightarrow X_n = 0 ,  \mathscr{R}(s) = 1 \\
	X_n < t_2 \rightarrow X_n = 0 ,  \mathscr{R}(s) = 0
\end{eqnarray}
where $\mathscr{R}(s)$ is the output of random variable $\mathscr{R}$ from the proposed method MRE-IPI. Higher threshold levels will increase the number of bits we need from IPI to generate one MRE-IPI random bit. However, lower $t$ values may not provide a good randomness quality. Based on the experiments that we have done, the best value for threshold level is $t_1=t_2=3$. 

\section{Analysis and Results}
In order to examine the randomness of MRE-IPI algorithm we used two approaches. In the first step, we conducted an entropy analysis on the result of MRE-IPI to examine the \textit{Robustness} and \textit{Permanence} of MRE-IPI. We did not test the condition of \textit{Uniqueness}, since in previous section, we showed that the dataset (number 2) satisfies this condition. The second step of analysis is to use random number test suites to check the quality of developed random numbers. To this end, we have used well known benchmarks such as NIST STS and Dieharder randomness test suites.

\subsection{Robustness}
As mentioned before, we used all the information secrecy measures to examine the \textit{Robustness} of MRE-IPI. There are three sets to compare: train dataset which the conversion to stochastic process has been done using this, test dataset which we did not use them through the development of MRE-IPI and both of them together as one dataset.

\begin{figure}
\centering
\includegraphics[width=\columnwidth]{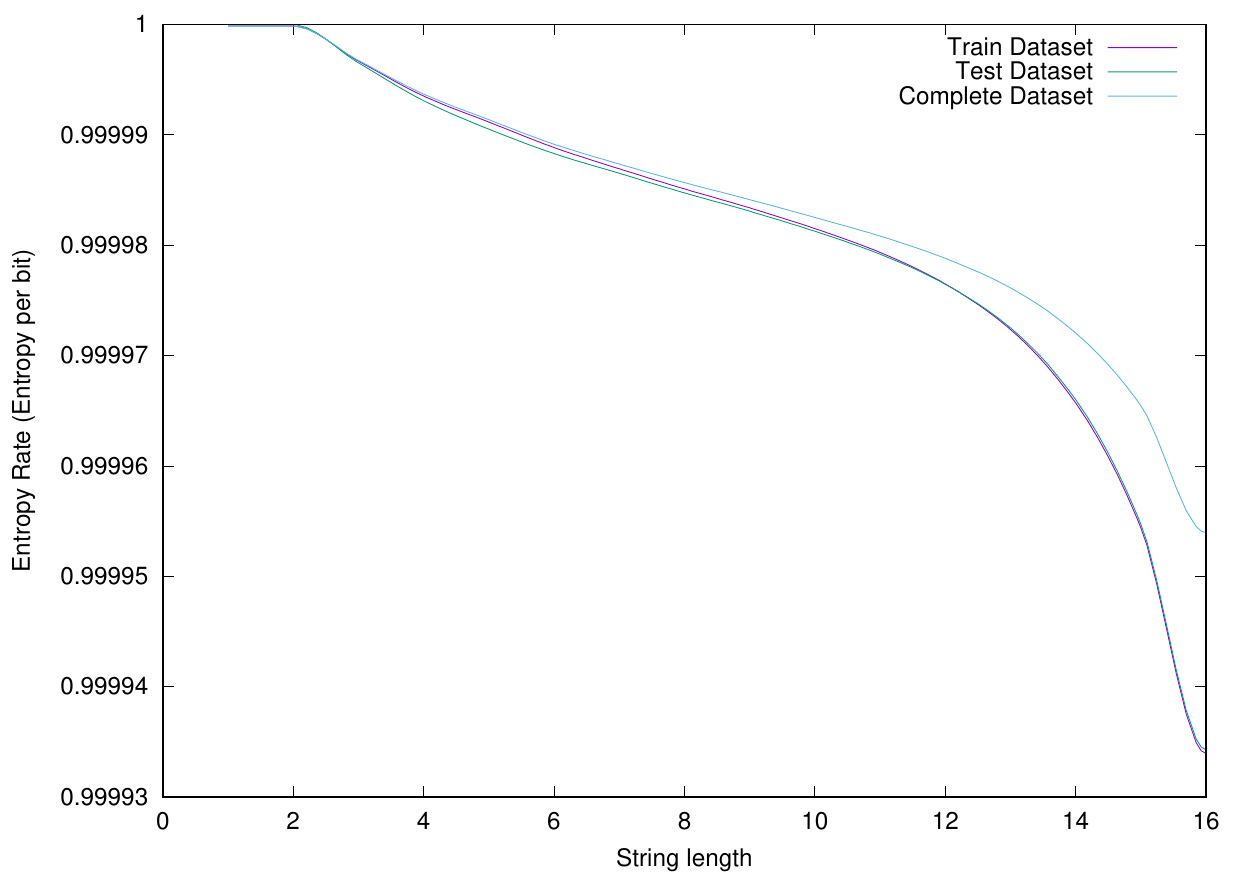}
\caption{Shannon Entropy of MRE-IPI datasets in various string lengths}
\label{fig:entropyshannonMRE}
\end{figure}

As shown in Fig.~\ref{fig:entropyshannonMRE} Shannon Entropy of MRE-IPI for up to string length of 16 is as high as 0.9999. Although, by increasing the size of the string to 16, the entropy rate drops from 1 to 0.99993, the difference is negligible. In total, MRE-IPI shows very high Shannon Entropy, and thus provides high quality randomness in \textit{Perfect Secrecy} assumption. 

\begin{figure}
\centering
\includegraphics[width=\columnwidth]{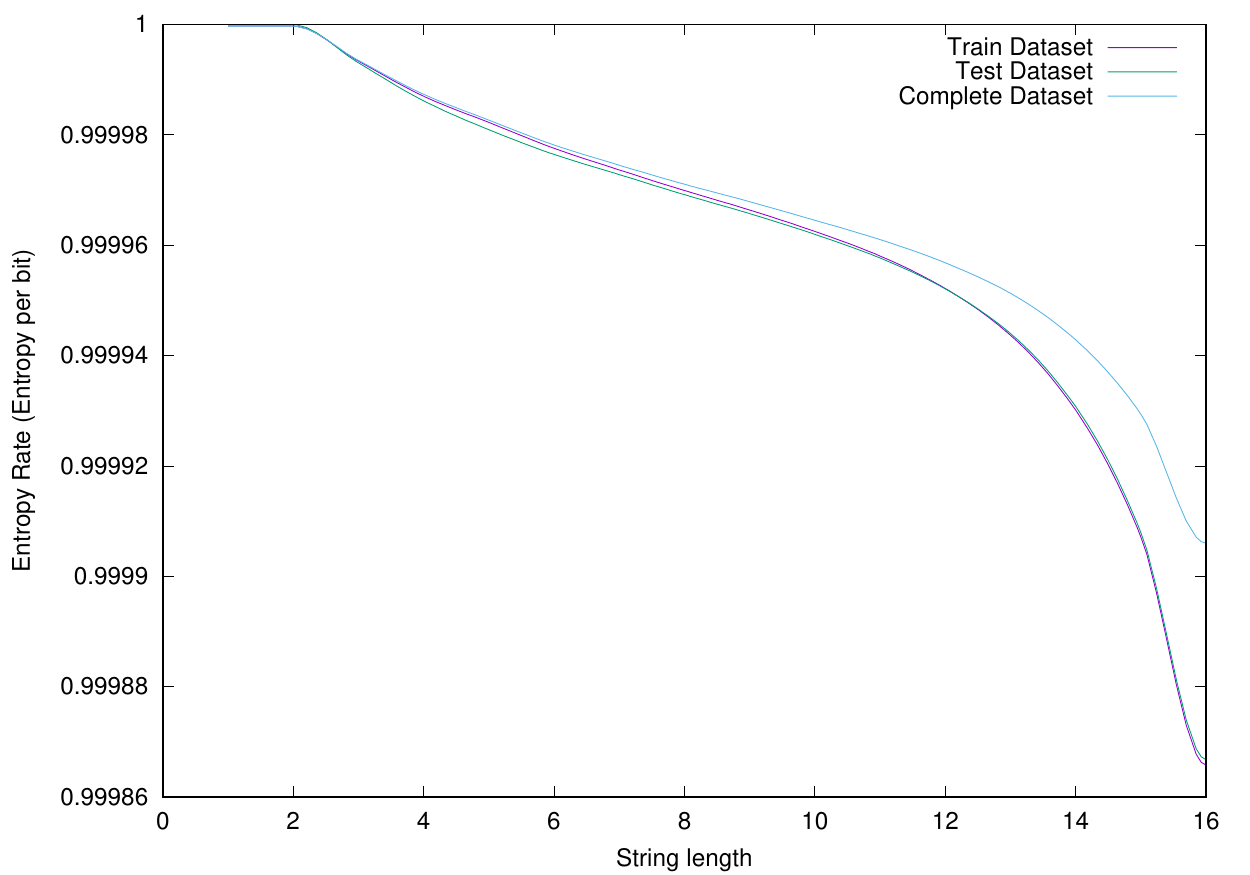}
\caption{Collision Entropy MRE-IPI datasets in various string lengths}
\label{fig:entropycollisionMRE}
\end{figure}

\begin{figure}
\centering
\includegraphics[width=\columnwidth]{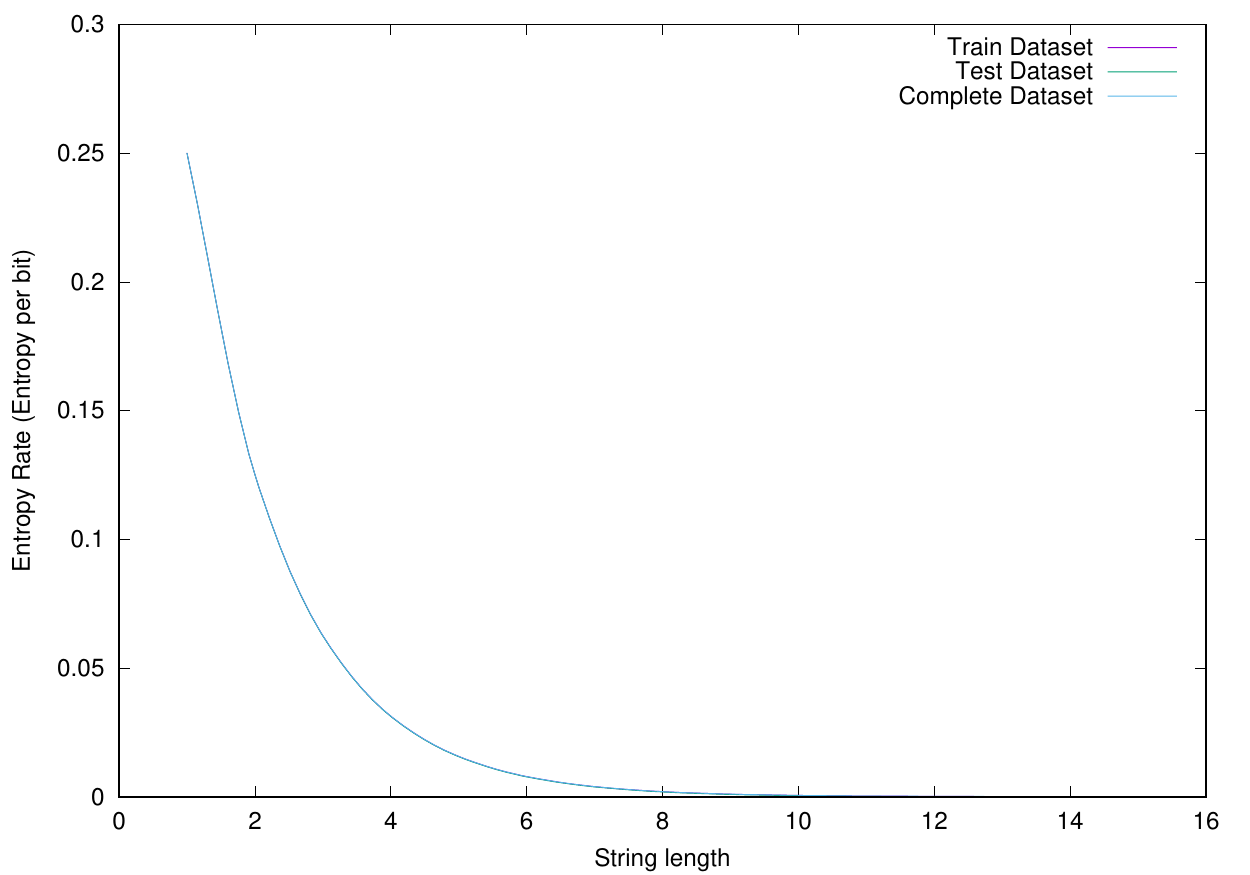}
\caption{Guessing Entropy of MRE-IPI datasets in various string lengths}
\label{fig:entropyguessMRE}
\end{figure}

Fig.~\ref{fig:entropycollisionMRE} shows the Collision Entropy of MRE-IPI. As shown, the entropy rate even up to string lengths of 16 for training, testing and complete datasets is higher than 0.99986. Fig.~\ref{fig:entropyguessMRE} presents the Guessing Entropy analysis of MRE-IPI. As shown, for string lengths higher than 6, the Guess Entropy (subtracted from 0.5) is less than 0.01 and for string length of 16 for all datasets are less than $1e^{-03}$. Thus, MRE-IPI is able to provide the the \textit{Robustness} in {Conditional Secrecy} scenario.

\begin{figure}
\centering
\includegraphics[width=\columnwidth]{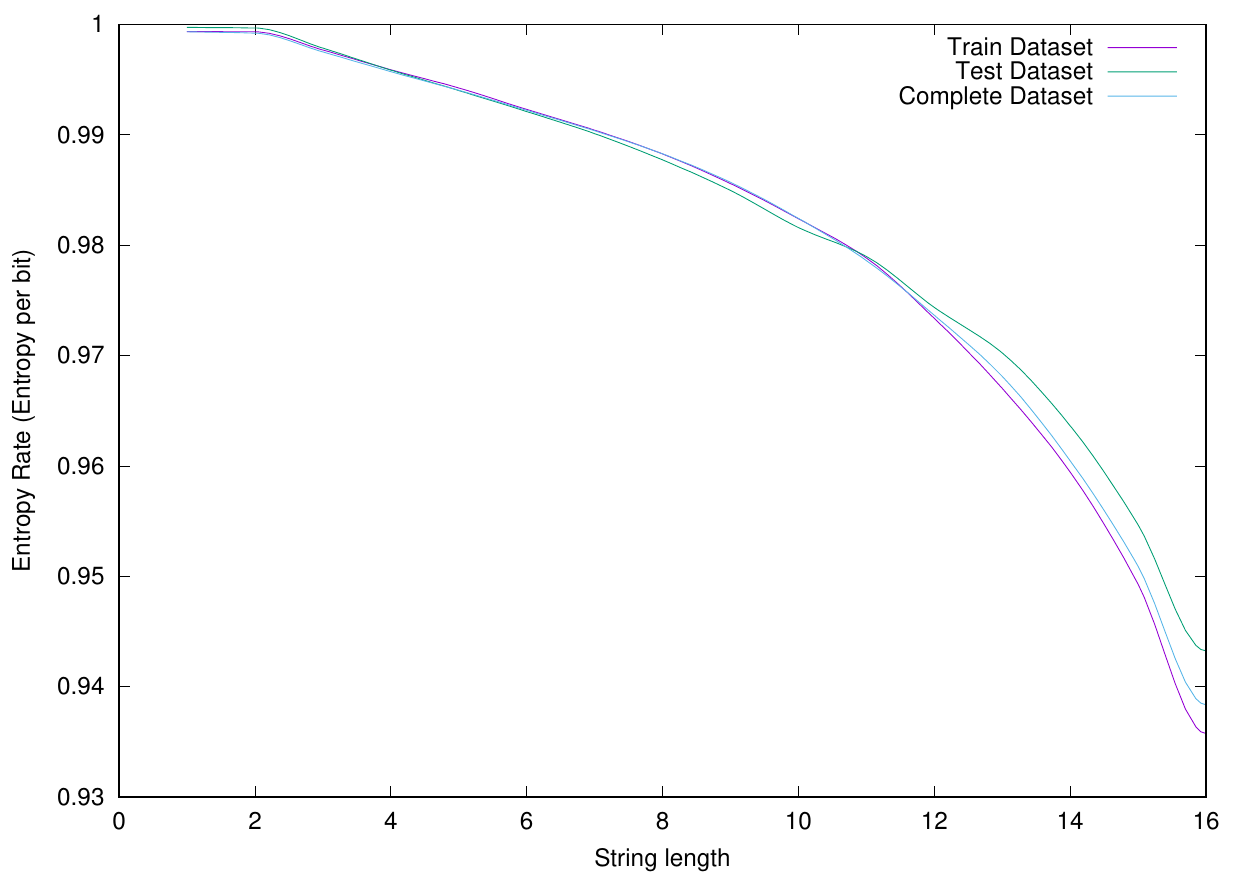}
\caption{Min-Entropy of MRE-IPI datasets in various string lengths}
\label{fig:entropyminMRE}
\end{figure}

For Min-Entropy (fig.~\ref{fig:entropyminMRE}), the values for train, test and complete datasets are almost equal. All the datasets, with the string length of 16, have Min-Entropy rate higher than 0.930. This shows a great increase in Min-Entropy of IPI after using MRE-IPI algorithm. Min-Entropy is a measure for \textit{Unconditional Secrecy} and MRE-IPI shows \textit{Robust} randomness in the worse case scenario (\textit{Unconditional Secrecy}).

\begin{figure}
\centering
\includegraphics[width=\columnwidth]{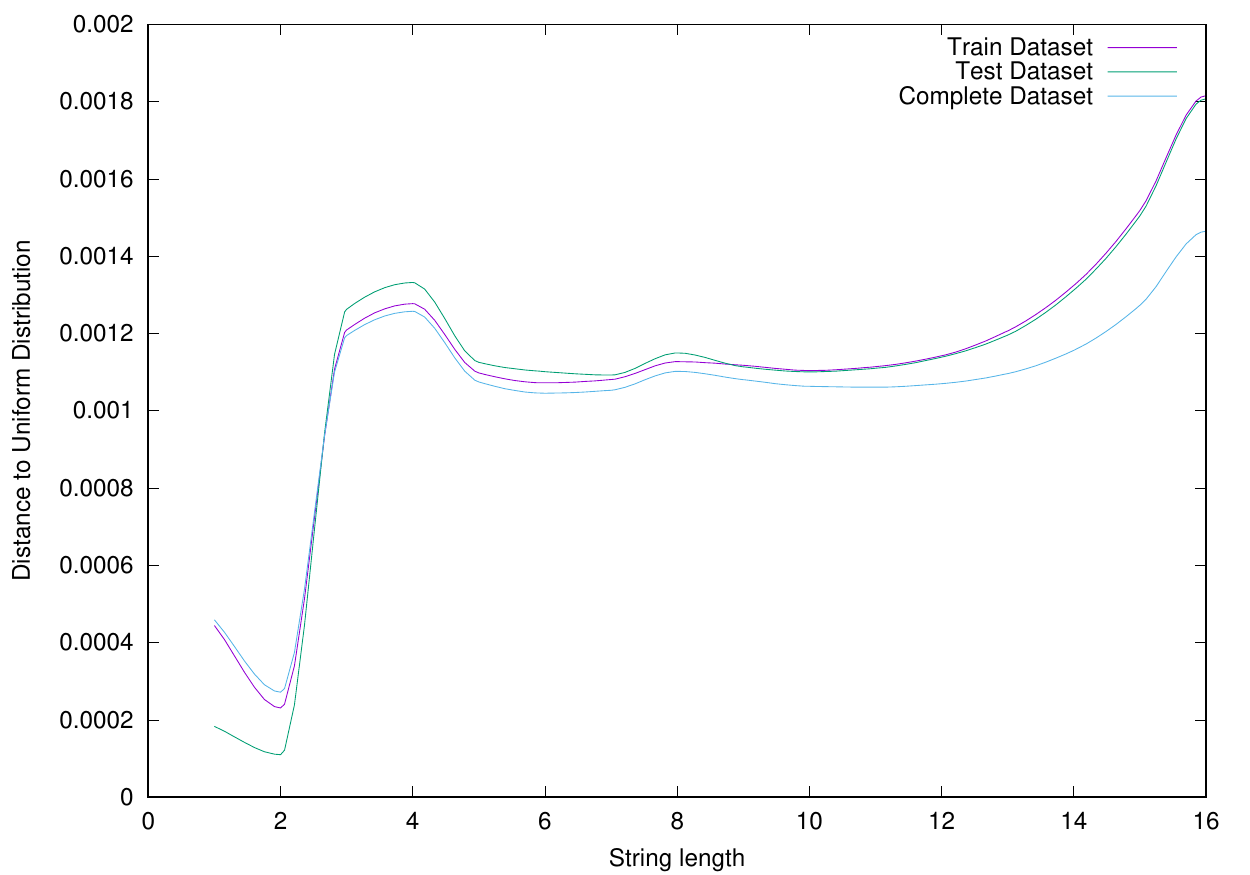}
\caption{Uniformity check for MRE-IPI datasets in various string lengths}
\label{fig:pboundMRE}
\end{figure}
 
Fig.~\ref{fig:pboundMRE} shows the uniformity check of the distribution of MRE-IPI databases. This is the \textit{Probabilistic Bound} one of the information secrecy measures. For all datasets in MRE-IPI (train, test and complete), the values are less than $0.02$ for all string lengths and thus, MRE-IPI shows almost uniform distribution. In total, MRE-IPI shows that it has the \textit{Robustness} condition of a strong randomness source.

\subsection{Permanence}  
\begin{figure}
\centering
\includegraphics[width=\columnwidth]{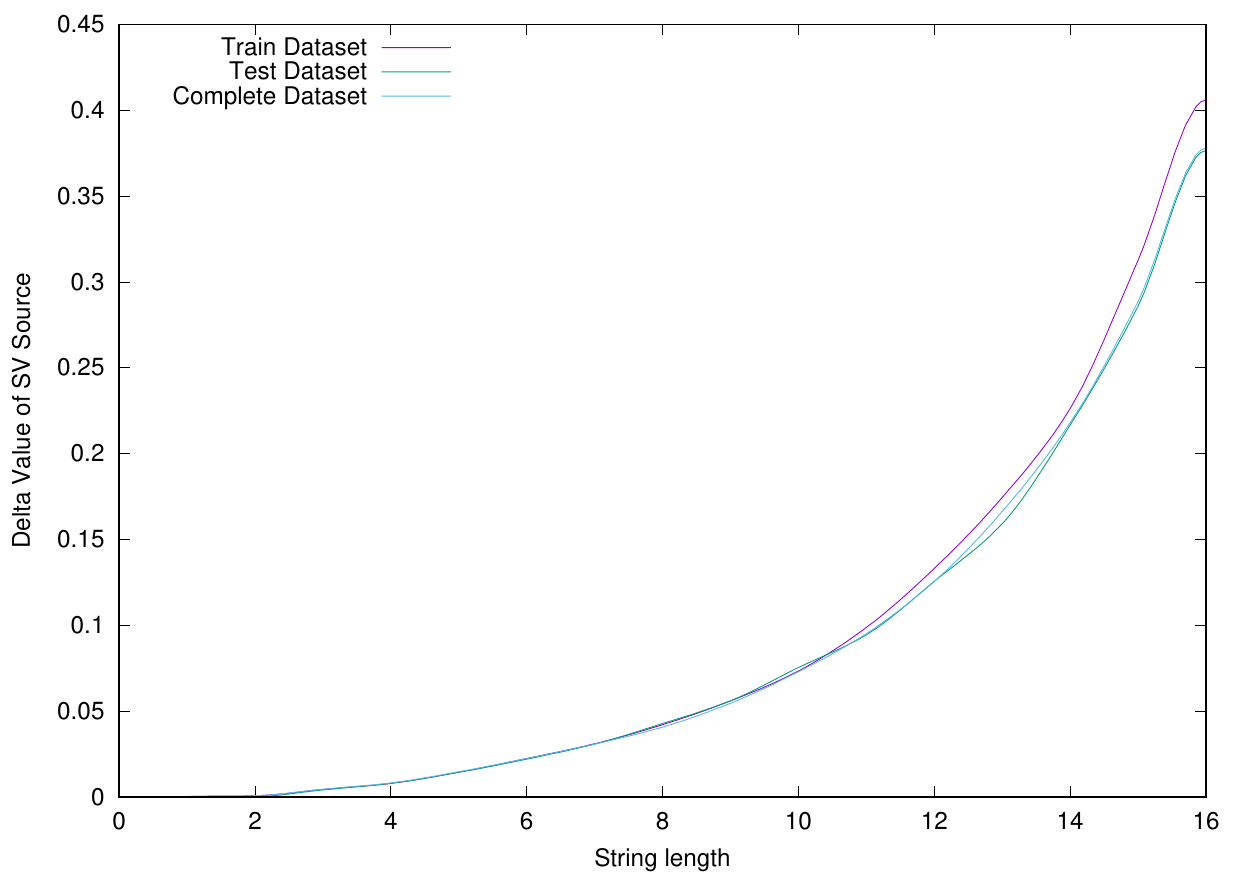}
\caption{$\delta$ value of SV-source for various string lengths of MRE-IPI}
\label{fig:deltaMRE}
\end{figure}

The next step is to check the \textit{Permanence} condition of MRE-IPI using the $\delta$ value of SV-source to see how much the MRE-IPI depends to its history. As shown in Fig.~\ref{fig:deltaMRE}, MRE-IPI has a great reduction in dependency compared to the IPI value itself. Up to string length of 10, $\delta$ is less than 0.1, and $\delta$ increases for string lengths 16 to a max value of 0.4. This is a significant improvement compare to IPI value in which $\delta$ for string length 16 is almost 1. This result shows that MRE-IPI has much lower dependency to its history compared to original IPI. Thus, it can provide \textit{Permanence} condition to some extent. In this situation, MRE-IPI is not recommended to be used as the secret key for a session, however, it could be used as a one-time-pad to exchange the session key such as \cite{Venkatasubramanian:2010bwb}. 

\subsection{Test Suites Analysis}
To evaluate a random extraction algorithm, a number of statistical tests have been proposed. A combination of randomness test algorithms is called a randomness suite or a battery of tests. The most used batteries of tests in randomness assessment are: NIST STS \cite{Rukhin:2000ee} and Dieharder \cite{Brown:2013wb}. A randomness suite usually needs a large set of random numbers for its tests. For instance, in NIST STS Linear test, the length of a sequence of bits must be greater than $10^6$ and at least $100$ sequences are needed to get a reliable result from the test. This is more than 3,000,000 of 32-bit integer numbers. Using the collected IPI dataset, in this research, we evaluated MRE-IPI compared to the recent IPI randomness extraction algorithms by 2-dimensional scatter plot and two batteries of tests: NIST STS, and Dieharder.

\subsubsection{Two Dimensional Scatter Plot}
\begin{figure*}
\centering
	\begin{subfigure}{0.33\textwidth}
        \centering
        \includegraphics[width = \textwidth]{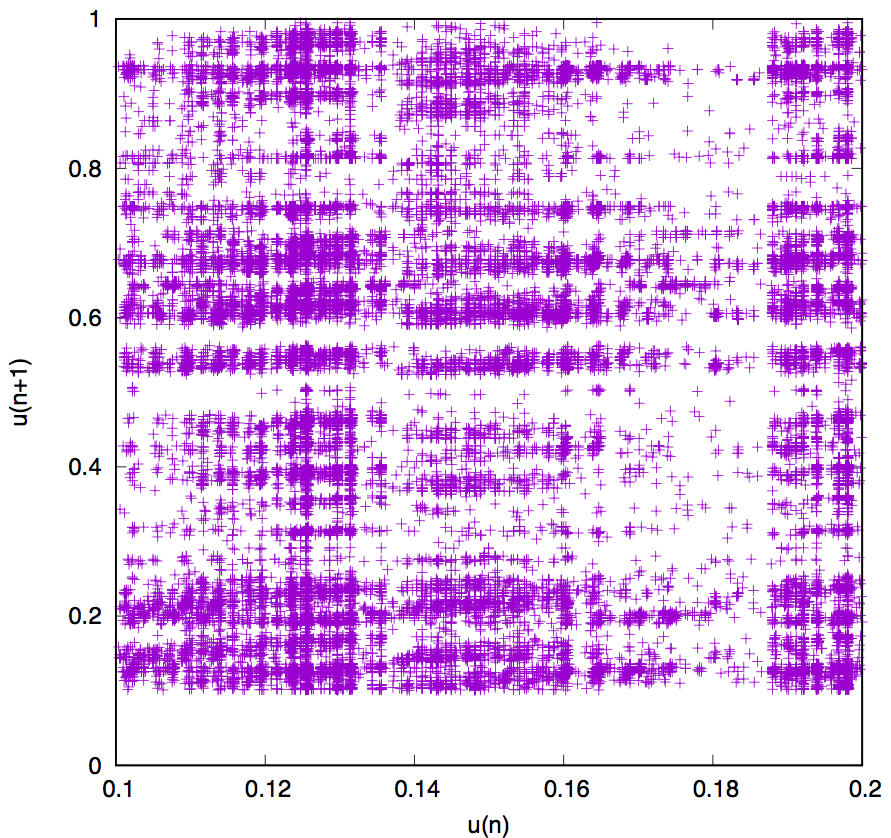}
        \caption{\cite{TianHong:2011ez}}
    \end{subfigure}%
	\begin{subfigure}{0.33\textwidth}
        \centering
        \includegraphics[width = \textwidth]{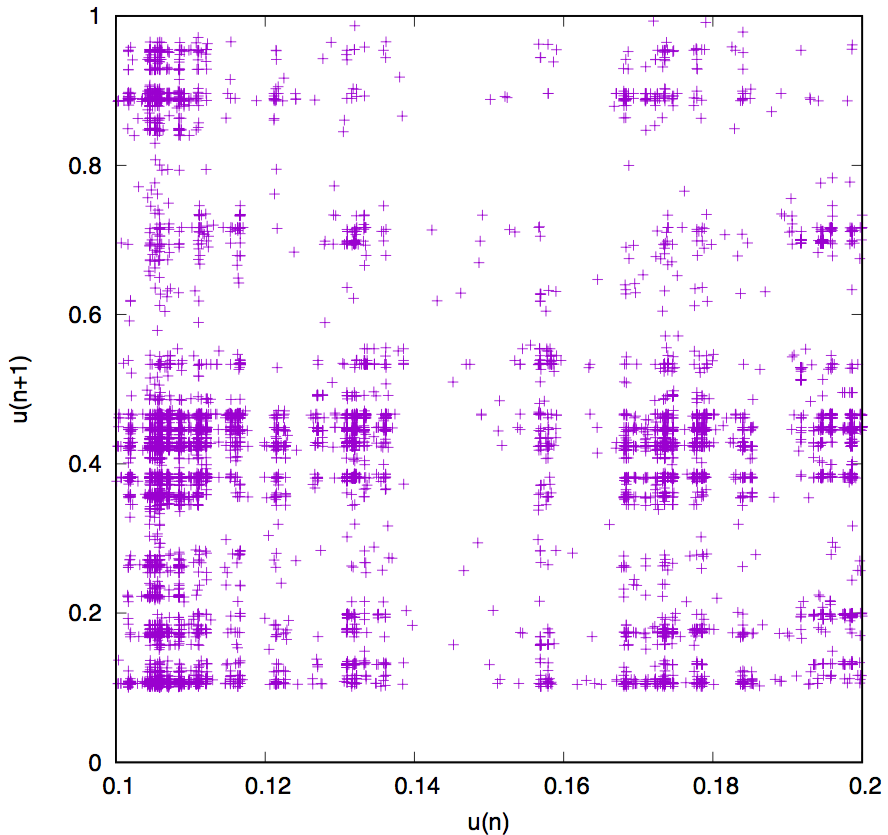}
        \caption{\cite{Bao:2012fw}}
    \end{subfigure}%
	\begin{subfigure}{0.33\textwidth}
        \centering
        \includegraphics[width = \textwidth]{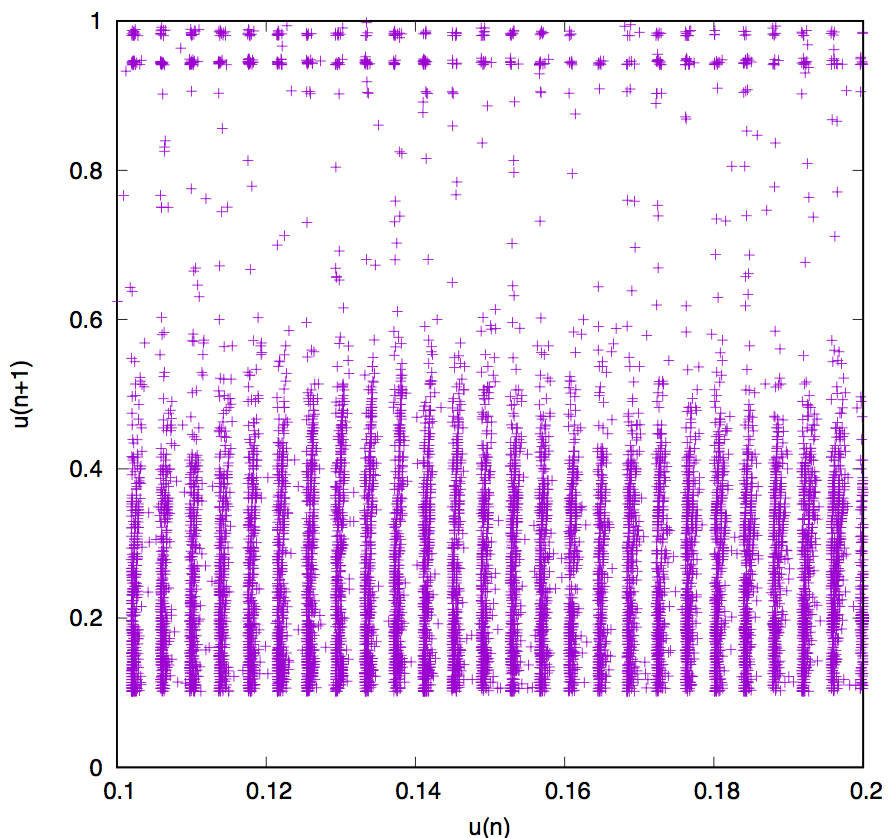}
        \caption{\cite{Cho:2012fp}}
    \end{subfigure}%
    \\
	\begin{subfigure}{0.33\textwidth}
        \centering
        \includegraphics[width = \textwidth]{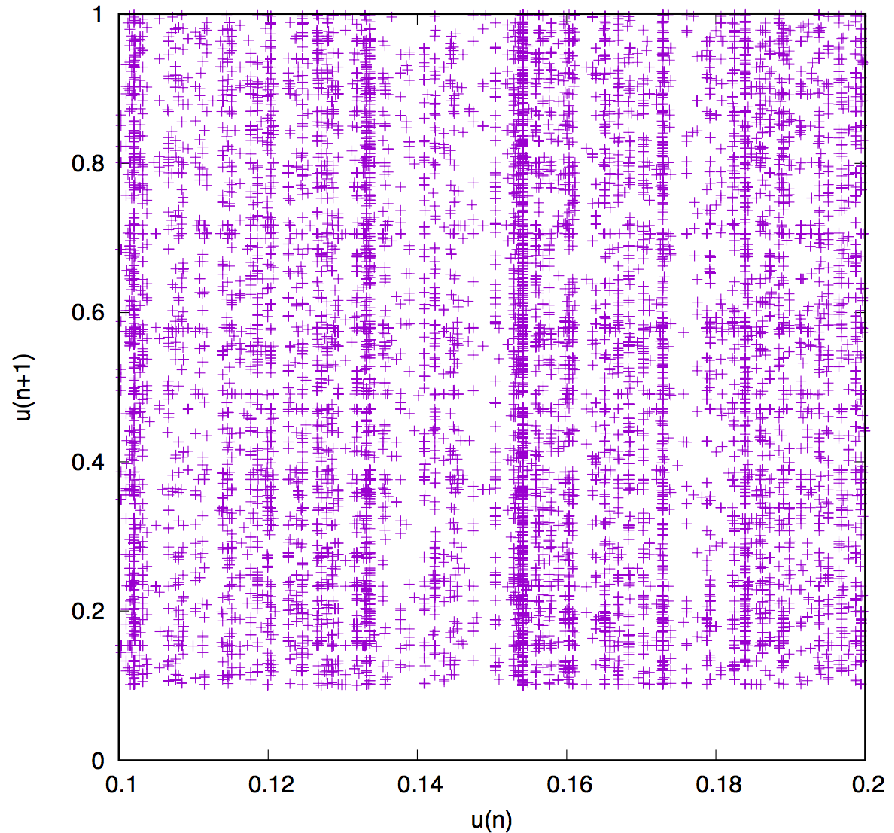}
        \caption{\cite{Hu:2013hs}}
    \end{subfigure}%
	\begin{subfigure}{0.33\textwidth}
        \centering
        \includegraphics[width = \textwidth]{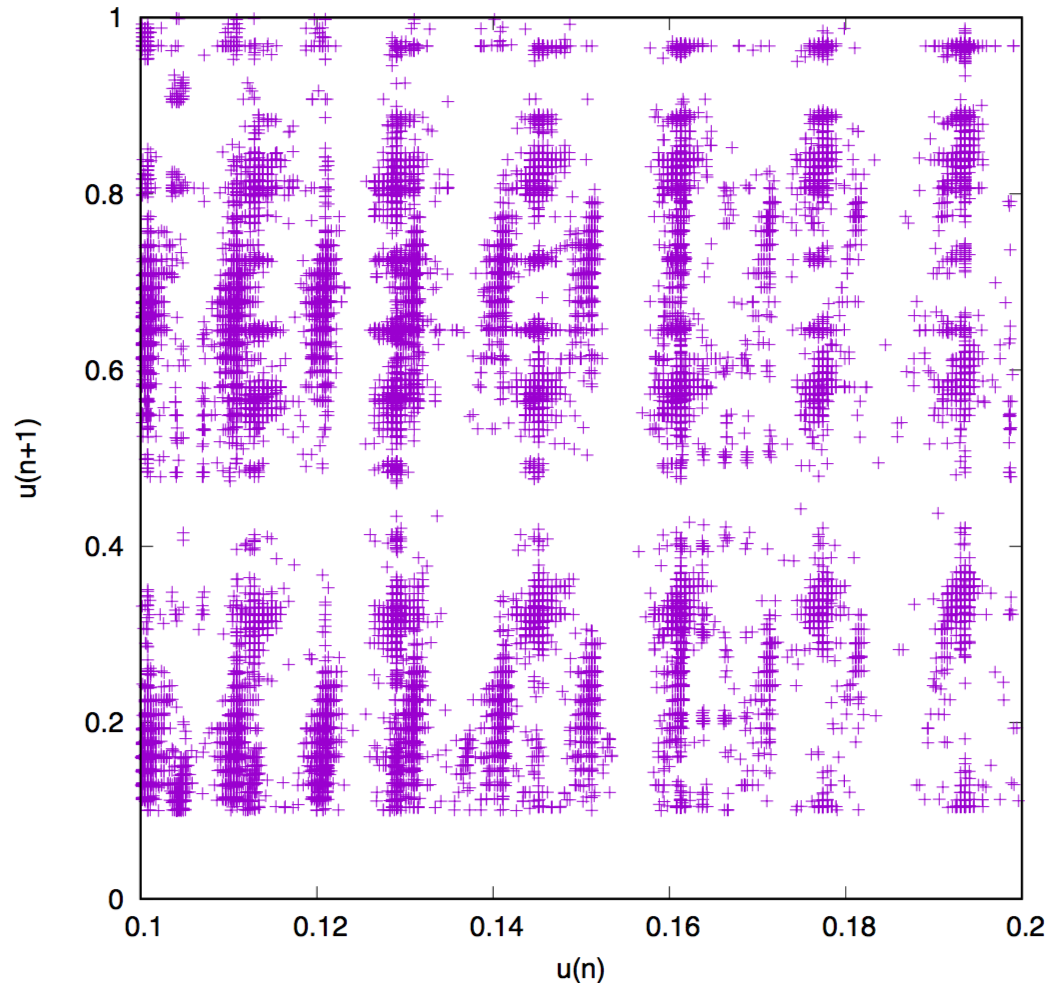}
        \caption{\cite{Altop:2015ej}}
    \end{subfigure}%
	\begin{subfigure}{0.33\textwidth}
        \centering
        \includegraphics[width = \textwidth]{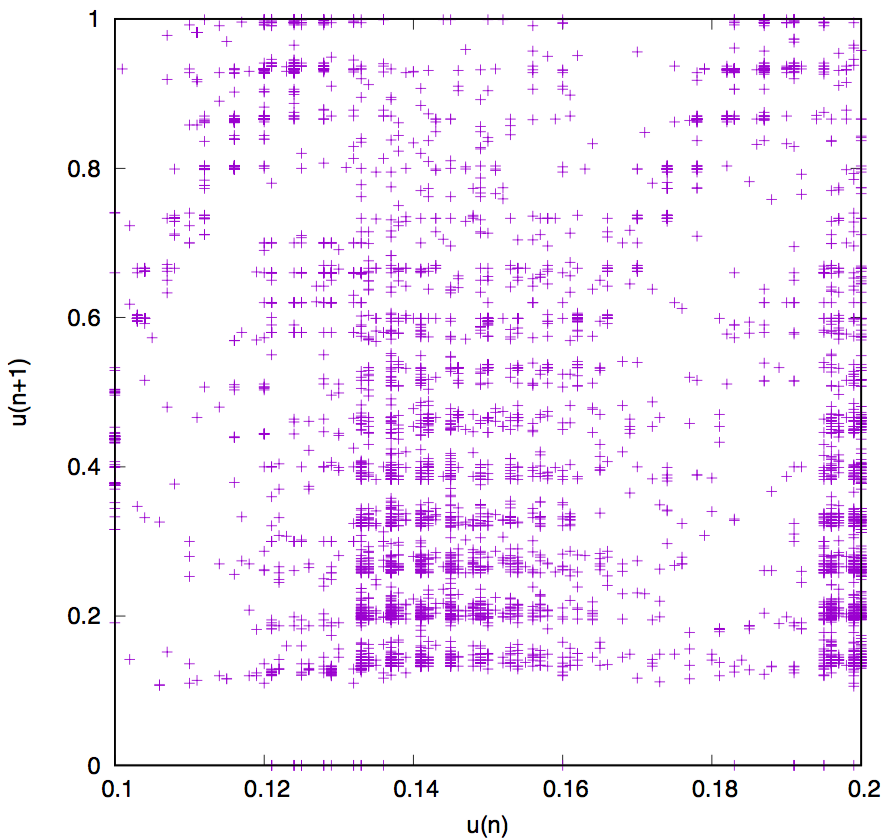}
        \caption{\cite{Jammali:2015ii}}
    \end{subfigure}%
    \\
	\begin{subfigure}{0.33\textwidth}
        \centering
        \includegraphics[width = \textwidth]{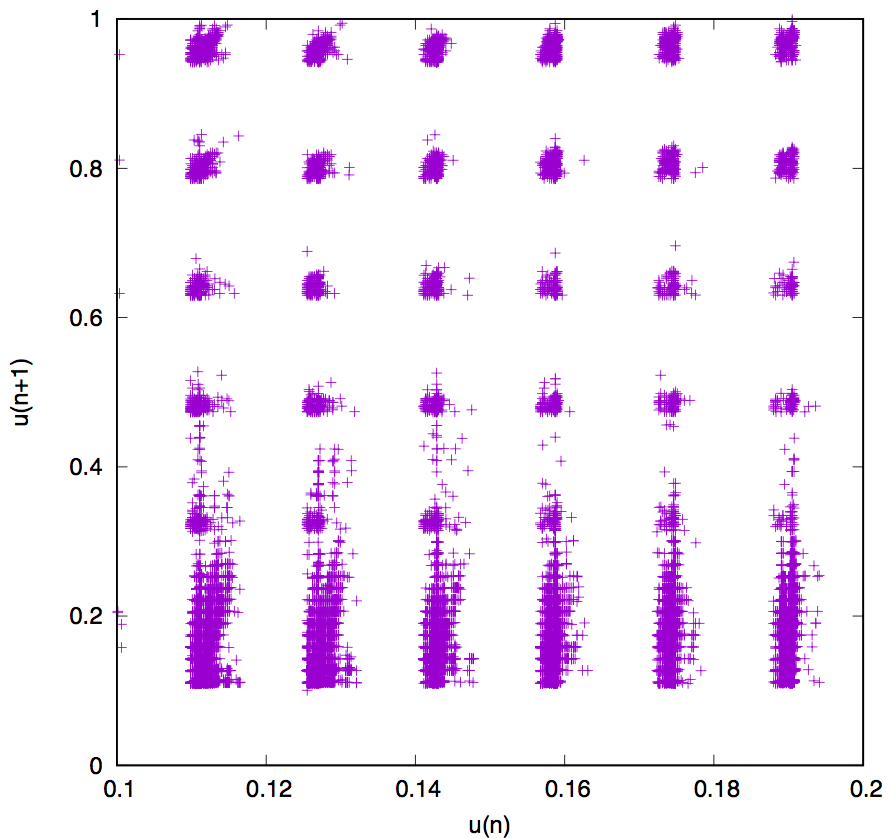}
        \caption{\cite{Peter:2016dv}}
    \end{subfigure}%
	\begin{subfigure}{0.33\textwidth}
        \centering
        \includegraphics[width = \textwidth]{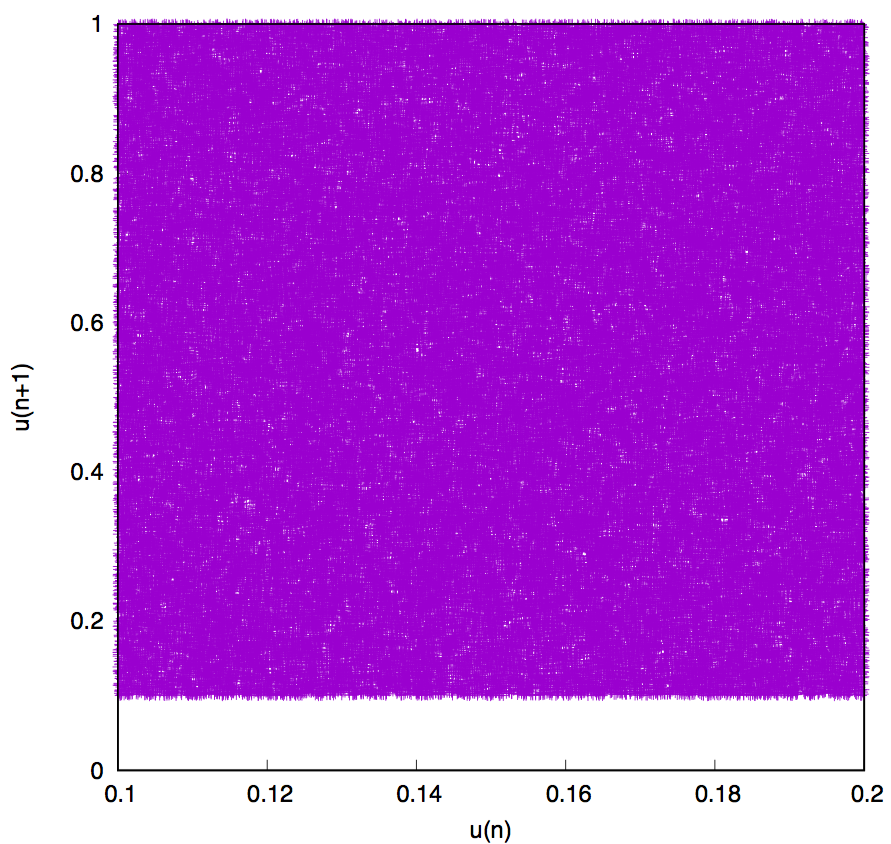}
        \caption{MRE-IPI}
    \end{subfigure}%
\caption{Uniformity check from 2D scatter plot (TestU01) comparing our result (h) with algorithms proposed in selected works (a-g)}
\label{fig:stsscatter}
\end{figure*}

The first step to examine the strength of randomness extractor algorithms is to produce a 2D scatter plots of $N$ points obtained from them. The $N$ points are generated in the t-dimensional unit hypercube $[0,1]^t$, either by taking vectors of $t$ successive output values from the extractor algorithm, or by taking $t$ non-successive values at pre-specified lags. The output of this 2D scatter plot should demonstrate a plane area without any specific pattern. To implement this, randomness statistical test suite with name TestU01 \cite{LEcuyer:2007ki} is used and the output is presented in Fig.~\ref{fig:stsscatter}. As shown, all the current randomness extractor algorithms left a pattern in their figure. However, MRE-IPI produces an almost uniform distribution of numbers without any recognisable pattern.

\subsubsection{NIST STS Test Suite}
NIST STS is the first statistical suite for randomness test several of them consisting multiple sub-tests. In total, there are 190 tests and sub-tests for NIST STS. In order to test the IPI randomness extractor algorithms, two series of examinations are taken place. In the first series, the binary dataset of random numbers is considered as the sequences with length 10,000 (10k) bits. In the second series of tests, 1,000,000 (1M) bits create a sequence to be tested in NIST STS. 

The reason is that in NIST STS with longer sequences, there will be more confident in the results. Moreover, there are some tests which cannot work with short bit sequences such as \textit{Random Excursions Test} and \textit{Random Excursions Variant Test}. However, if we increase the length of sequences, the number of available sequences will be reduced which decreases the confident in the results as well. In order to have a better picture of the quality of the results of NIST STS, we used both short (10k bits) and long (1M bits) sequences in the experiments. 

We showed in Table~\ref{tbl:niststs} the results of NIST STS Test Suite on the works \cite{TianHong:2011ez,Bao:2012fw,Choto:2012is,Hu:2013hs,Altop:2015ej,Jammali:2015ii,Peter:2016dv} compared with MRE-IPI. The algorithms proposed in \cite{Bao:2012fw} and \cite{Miao:2012cb} are very similar and thus one column for their results is used for them. \cite{Peter:2016dv}, \cite{Zheng:2015bx} and \cite{Zheng:2014uh} also proposed very similar extraction algorithms and thus one column is presenting the results for them. As shown in the table, \cite{Jammali:2015ii,Peter:2016dv} are the closest works to MRE-IPI by 55\% and 57\% rate of success in passing tests for sequence length 10k, while MRE-IPI has passed 80\% of the NIST STS with the same sequence length. Moving to sequences with length 1M bits, the highest passing rate of 12\% tests belongs to \cite{Hu:2013hs}, while MRE-IPI has the pass rate of 86\% for NIST STS tests. That clearly shows the robustness of the extraction method which with various length of sequences in testing, provides high quality results.

\begin{sidewaystable}
\centering
\scriptsize
\caption{NIST STS Test Suite results for MRE-IPI and other IPI randomness extraction methods}
\begin{tabular}{|p{3cm} | c | c | c | c | c | c | c | c | c | c | c | c | c | c | c | c | c | c |}
\hline
Test 	&	No. of	&	\multicolumn{2}{|c|}{\cite{TianHong:2011ez}}		&	\multicolumn{2}{|c|}{\cite{Bao:2012fw}}	&	\multicolumn{2}{|c|}{\cite{Choto:2012is}}	&	\multicolumn{2}{|c|}{\cite{Hu:2013hs}}	&	\multicolumn{2}{|c|}{\cite{Altop:2015ej}}	&	\multicolumn{2}{|c|}{\cite{Jammali:2015ii}}	&	\multicolumn{2}{|c|}{\cite{Peter:2016dv}}	&	\multicolumn{2}{|c|}{MRE-IPI}	\\ \cline{3-18}
Name	&	tests	&	10k	&	1M	&	10k	&	1M	&	10k	&	1M	&	10k	&	1M	&	10k	&	1M	&	10k	&	1M	&	10k	&	1M	&	10k	&	1M	\\ \hline
Frequency (Monobit)	&	1	&	0	&	0	&	0	&	0	&	0	&	0	&	0	&	0	&	0	&	0	&	0	&	0	&	0	&	0	&	1	&	0	\\ \hline
Block Frequency	&	1	&	0	&	0	&	0	&	0	&	0	&	0	&	0	&	0	&	0	&	0	&	0	&	0	&	0	&	0	&	0	&	0	\\ \hline
Cumulative Sums	&	2	&	0	&	0	&	0	&	0	&	0	&	0	&	0	&	0	&	0	&	0	&	0	&	0	&	0	&	0	&	1	&	0	\\ \hline
Runs	&	1	&	0	&	0	&	0	&	0	&	0	&	0	&	0	&	0	&	0	&	0	&	0	&	0	&	0	&	0	&	1	&	1	\\ \hline
Longest Run	&	1	&	0	&	0	&	0	&	0	&	0	&	0	&	0	&	0	&	0	&	0	&	0	&	0	&	0	&	0	&	1	&	1	\\ \hline
Binary Matrix Rank	&	1	&	0	&	0	&	0	&	0	&	0	&	0	&	0	&	0	&	0	&	0	&	0	&	0	&	0	&	0	&	1	&	1	\\ \hline
Discrete Fourier Transform	&	1	&	0	&	0	&	0	&	0	&	0	&	0	&	0	&	0	&	0	&	0	&	0	&	0	&	0	&	0	&	1	&	1	\\ \hline
Non-overlapping Template Matching 	&	150	&	5	&	0	&	65	&	0	&	0	&	0	&	19	&	0	&	68	&	0	&	105	&	0	&	109	&	0	&	143	&	131	\\ \hline
Overlapping Template Matching 	&	1	&	0	&	0	&	0	&	0	&	0	&	0	&	0	&	0	&	0	&	0	&	0	&	0	&	0	&	0	&	1	&	0	\\ \hline
Maurer’s Universal	&	1	&	0	&	0	&	0	&	0	&	0	&	0	&	0	&	0	&	0	&	0	&	0	&	0	&	0	&	0	&	0	&	0	\\ \hline
Approximate Entropy	&	1	&	0	&	0	&	0	&	0	&	0	&	0	&	0	&	0	&	0	&	0	&	0	&	0	&	0	&	0	&	0	&	0	\\ \hline
Random Excursions	&	8	&	-	&	0	&	-	&	0	&	-	&	0	&	-	&	5	&	-	&	0	&	-	&	0	&	-	&	0	&	-	&	8	\\ \hline
Random Excursions Variant	&	18	&	-	&	11	&	-	&	13	&	-	&	18	&	-	&	18	&	-	&	0	&	-	&	4	&	-	&	1	&	-	&	18	\\ \hline
Serial	&	2	&	0	&	0	&	0	&	0	&	0	&	0	&	0	&	0	&	0	&	0	&	0	&	0	&	0	&	0	&	2	&	2	\\ \hline
Linear Complexity	&	1	&	0	&	1	&	0	&	0	&	0	&	1	&	0	&	0	&	0	&	0	&	0	&	0	&	0	&	0	&	0	&	1	\\ \hline
Total	&	190	&	5	&	12	&	65	&	13	&	0	&	19	&	19	&	23	&	68	&	0	&	105	&	4	&	109	&	1	&	152	&	164	\\ \hline
Percentage	&	1.00	&	0.03	&	0.06	&	0.34	&	0.07	&	0.00	&	0.10	&	0.10	&	0.12	&	0.36	&	0.00	&	0.55	&	0.02	&	0.57	&	0.01	&	0.80	&	0.86	\\ \hline
\end{tabular}
\label{tbl:niststs}
\end{sidewaystable}

In addition to success rate in passing the tests, checking the proportion of sequences passing a test is another measure to examine the quality of a randomness extractor. The confidence interval ($ci$) for proportion of the binary sequences which passed the tests is calculated from Eq.~\ref{eq:ci}, where $\hat{p}$ is $1-\alpha$ and $m$ is the number of sequences.

\begin{equation}
ci = \hat{p} \pm 3 \sqrt{\frac{\hat{3}(1-\hat{p})}{m}}
\label{eq:ci}
\end{equation}

\begin{figure}
\centering
\includegraphics[width=1.0\columnwidth]{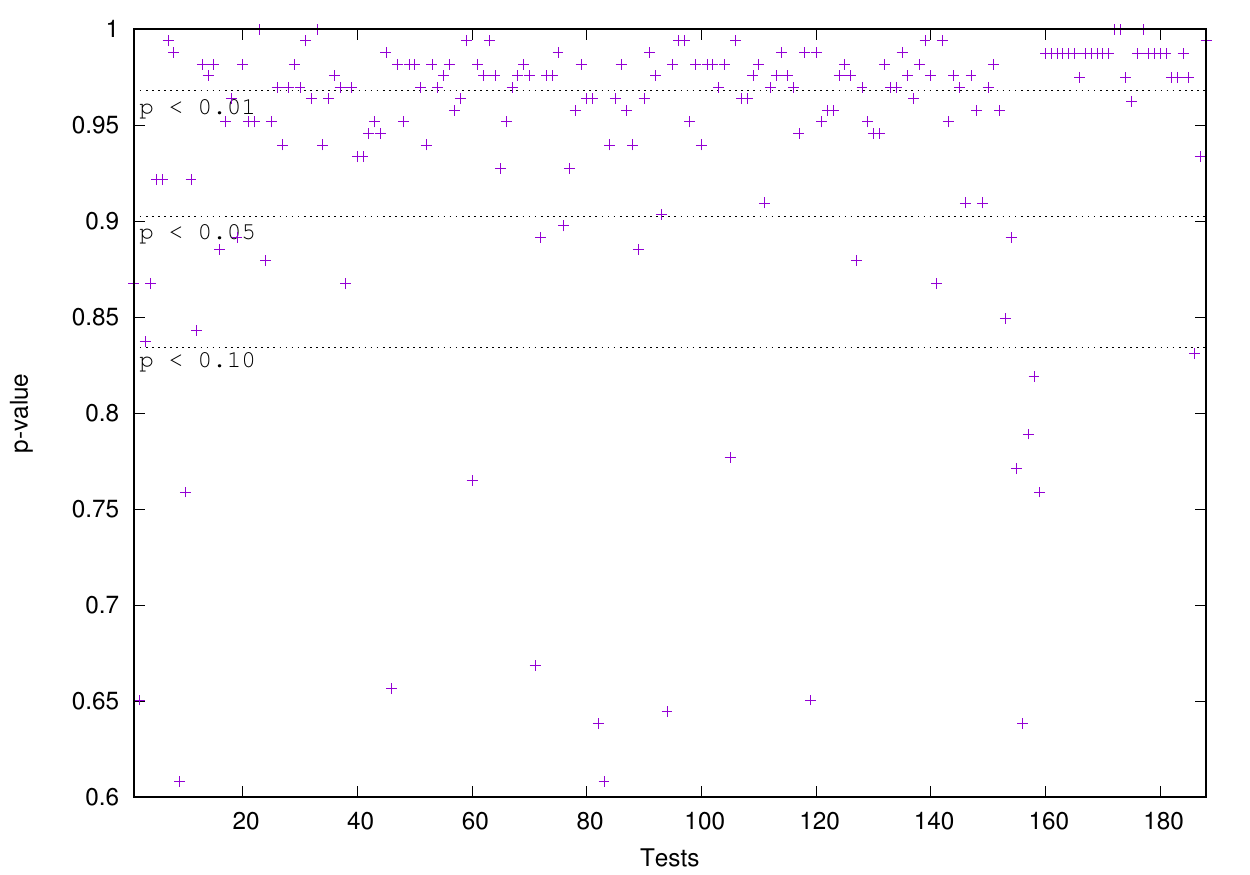}
\caption{Scatter plot of the p-values for NIST STS tests}
\label{fig:sts2scatter}
\end{figure}

The $alpha$ value and the percentage of sequences which passed that confidence interval for MRE-IPI and other methods in literature are presented in Table~\ref{tbl:ci}. Base on this table, with 95\% confidence more than 82\% of sequences have passed the test for MRE-IPI, which is an improvement over the closest result belongs to \cite{Peter:2016dv} with 71.28\% pass rate of sequences with 95\% confidence. Figure~\ref{fig:sts2scatter} shows the scatter plot of p-values for NIST STS tests. Although these figures suggest slight improvement compared to other methods, we will show in the next section that our solution is a much stronger randomness generator in practice.  

\begin{table}
\centering
\small
\caption{Percentage of sequences which passed the tests based on various $\alpha$ values}
\begin{tabular}{|l | c | c | c | c | c | c | }
\hline
Extraction	&	\multicolumn{3}{|c|}{Seq size = 10,000}	\\ \cline{2-4}
Methods	&	$\alpha=0.10$	&	$\alpha=0.05$	&	$\alpha=0.01$	 \\ \hline
\cite{TianHong:2011ez}	&	35.64	&	23.93	&	9.04	 \\ \hline
\cite{Bao:2012fw}		&	58.51	&	54.26	&	45.74	 \\ \hline
\cite{Cho:2012fp}		&	47.87	&	27.66	&	0.00	\\ \hline
\cite{Hu:2013hs}		&	51.60	&	36.70	&	13.83	 \\ \hline
\cite{Altop:2015ej}		&	65.43	&	61.17	&	48.40	 \\ \hline
\cite{Jammali:2015ii}	&	64.89	&	64.89	&	47.87	 \\ \hline
\cite{Peter:2016dv}		&	71.28	&	70.21	&	59.57 \\ \hline
MRE-IPI				&	85.11	&	85.11	&	76.60	\\ \hline
Extraction	&		\multicolumn{3}{|c|}{Seq size = 1,000,000}	\\ \cline{2-4}
Methods	&	 $\alpha=0.10$	&	$\alpha=0.05$	&	$\alpha=0.01$ \\ \hline
\cite{TianHong:2011ez}	&		10.64	&	7.45	&	1.06 \\ \hline
\cite{Bao:2012fw}		&		3.19	&	2.13	&	0.00 \\ \hline
\cite{Cho:2012fp}		&		10.11	&	10.11	&	10.11\\ \hline
\cite{Hu:2013hs}		&		8.51	&	8.51	&	8.51 \\ \hline
\cite{Altop:2015ej}		&	0.00	&	0.00	&	0.00 \\ \hline
\cite{Jammali:2015ii}	&	0.00	&	0.00	&	0.00 \\ \hline
\cite{Peter:2016dv}		&	0.53	&	0.53	&	0.53 \\ \hline
MRE-IPI				&	90.96	&	82.98	&	54.79\\ \hline
\end{tabular}
\label{tbl:ci}
\end{table}

\subsection{Dieharder Test Suite}
The final step to examine the quality of MRE-IPI is to use Dieharder randomness test suite. Dieharder consists of 31 tests  some include several sub-tests. In overall there are 114 tests and sub-tests. The output of Dieharder provides three ranks for each test including: PASSED, WEAK and FAILED. We implemented the recent randomness extraction algorithms from IPI (\cite{TianHong:2011ez,Bao:2012fw,Cho:2012fp,Hu:2013hs,Jammali:2015ii,Altop:2015ej,Peter:2016dv}), all of them have been FAILED in all the tests in Dieharder test suite. So, to evaluate the quality of MRE-IPI, we selected four standard random number generators. The first one is Microsoft Excel, the second one is Niederreiter, Borosh random number generator \cite{Niederreiter:2012bb}, the third one is Robert R. (Bob) Coveyou random number generator \cite{coveyou1969random} and finally the last one is AES random number generator (e.g. \cite{LEcuyer:2011wp}). For the last three random number generators, we used Dieharder randomness test suite for both generation and test. As shown in Table~\ref{tbl:diehard}, MRE-IPI completely outperforms Excel and \cite{Niederreiter:2012bb} random number generators. The quality of MRE-IPI is slightly better than \cite{coveyou1969random}, however, it is still far below AES. The reason for this could be the problem of \textit{Permanence} condition in MRE-IPI. Although it is much better than IPI itself, the independency to history still is not 100\% and perfect. Thus, MRE-IPI failed in the tests such as OPSO, OQSO, DNA and DAB DCT (Table~\ref{tbl:diehard}). This shows that although MRE-IPI has high quality in \textit{Robustness} and \textit{Uniqueness}, it may not be suitable to be used as a general cryptographic key. Because of not perfect \textit{Permanence} condition, our suggestion is to use additional tools such as hashing to convert MRE-IPI to a cryptographic key. This is maybe the case and we will evaluate this in future work. However, we showed already that the MRE-IPI is sufficiently good to be used in a cryptosystem. Nevertheless, due to the limited resources in IMDs, MRE-IPI can be used as a one-time pad for exchanging session key. For example, the gateway (outside the body) and IMD negotiated on a time and start to measure the IPI. Then, using MRE-IPI they create a one-time pad with the length same as the main key of the session. The gateway, produces the session key and encrypts it with a simple operation like XOR using the one-time pad and sends it to the IMD. Finally, IMD using its own one-time pad decrypts the received message and store the session key for encryption and decryption of the next messages. In this scheme the usage of MRE-IPI is reduced to a very few messages. 

\begin{table}
\centering
\small
\caption{Dieharder Test Suite results for proposed method and some conventional random number generators}
\begin{tabular}{|p{2.5cm} | c | c | c | c | c | }
\hline
Test Name	&	Excel	&	\cite{Niederreiter:2012bb}	&	\cite{coveyou1969random}	&	AES	&	MRE-IPI	\\ \hline
Birthdays 	&	0	&	1	&	1	&	1	&	1	\\ \hline
OPERM5 	&	0	&	1	&	1	&	1	&	1	\\ \hline
32x32 Binary Rank 	&	0	&	0	&	0	&	1	&	1	\\ \hline
6x8 Binary Rank 	&	0	&	0	&	0	&	1	&	0	\\ \hline
Bitstream 	&	0	&	0	&	0	&	1	&	0	\\ \hline
OPSO	&	0	&	0	&	0	&	1	&	0	\\ \hline
OQSO 	&	0	&	0	&	0	&	1	&	0	\\ \hline
DNA 	&	0	&	0	&	0	&	1	&	0	\\ \hline
Count the 1s (stream) 	&	0	&	0	&	0	&	1	&	0	\\ \hline
Count the 1s  (byte)	&	0	&	0	&	0	&	1	&	1	\\ \hline
Parking Lot 	&	0	&	1	&	1	&	1	&	1	\\ \hline
Minimum Distance (2d Circle) 	&	0	&	1	&	1	&	0	&	1	\\ \hline
3d Sphere (Minimum Distance) 	&	0	&	1	&	1	&	1	&	1	\\ \hline
Squeeze 	&	0	&	1	&	1	&	1	&	0	\\ \hline
Sums 	&	0	&	1	&	1	&	1	&	1	\\ \hline
Runs 	&	0	&	0	&	1	&	1	&	1	\\ \hline
Craps 	&	0	&	0	&	1	&	1	&	1	\\ \hline
Marsaglia and Tsang GCD 	&	0	&	0	&	0	&	1	&	1	\\ \hline
STS Monobit 	&	0	&	0	&	1	&	1	&	1	\\ \hline
STS Runs 	&	0	&	0	&	0	&	1	&	1	\\ \hline
STS Serial  (Generalized)	&	0	&	0	&	0	&	1	&	0	\\ \hline
RGB Bit Distribution 	&	0	&	0	&	0	&	1	&	0	\\ \hline
RGB Generalized Minimum Distance 	&	0	&	0	&	1	&	1	&	0	\\ \hline
RGB Permutations    	&	0	&	1	&	1	&	0	&	1	\\ \hline
RGB Lagged Sum 	&	0	&	0	&	0	&	1	&	0	\\ \hline
RGB Kolmogorov-Smirnov  	&	0	&	0	&	1	&	1	&	0	\\ \hline
Byte Distribution	&	0	&	0	&	0	&	1	&	0	\\ \hline
DAB DCT	&	0	&	0	&	0	&	1	&	0	\\ \hline
DAB Fill Tree 	&	0	&	1	&	0	&	1	&	0	\\ \hline
DAB Fill Tree 2 	&	0	&	0	&	0	&	1	&	0	\\ \hline
DAB Monobit 2 	&	0	&	0	&	0	&	1	&	0	\\ \hline
Total	&	0	&	9	&	13	&	29	&	14	\\ \hline
Percentage	&	0.00	&	0.29	&	0.42	&	0.94	&	0.45	\\ \hline
\end{tabular}
\label{tbl:diehard}
\end{table}

\section{Conclusion}
Although previous work has advocated the use of the physiological value in securing Body Sensor Network and access to IMDs, the application proposed for randomness extraction led to weak security and were not throughly evaluated. In this paper, using a sophisticated dataset with almost 900,000,000 IPI values, we measured the three condition of a strong random source: \textit{Uniqueness}, \textit{Robustness} and \textit{Permanence}. we showed that IPI value satisfies the condition of \textit{Uniqueness} when up to four least significant bits of it are being used as the source of randomness. IPI is a \textit{Robust} random source where the scenario in \textit{Perfect Secrecy} or \textit{Conditional Secrecy}. However, it is not able to provide the \textit{Robustness} in the worst case scenario or \textit{Unconditional Secrecy}. Moreover, we showed that IPI does not have the condition of \textit{Permanence} as it is highly depended to its history. Thus, we do not recommend extraction of random number from the IPI value. However, rather than using the value of IPI, we proposed using its trend. By converting the IPI trend to a Martingale Stochastic Process, we developed a random number extraction method named as MRE-IPI. Despite IPI, MRE-IPI provides \textit{Robustness} in all information secrecy measures, even in \textit{Unconditional Secrecy}. As it only uses the first two least significant bits of IPI, it also satisfies the condition of \textit{Uniqueness}. MRE-IPI has much lower dependency to its history compared to IPI and thus has better \textit{Permanence} condition. However, still it is not a completely independent from its history. The reason is that IPI itself is not a strong randomness source with highly dependency to the history. MRE-IPI was able to cure this dependency by looking at the trend of IPI rather than its value, which is not completely independent from the value of IPI. Furthermore, using randomness test suites such as NIST STS and Dieharder, we showed that MRE-IPI has an almost perfect score in NIST STS suite and could pass Dieharder with a quality as half as AES random number generator, while other IPI randomness extraction algorithms failed to pass the tests. We conclude that, as MRE-IPI has the properties of \textit{Robustness}, \textit{Uniqueness} and to some level the property of \textit{Permanence}, and as it is measurable in every person (\textit{Universality}) everywhere and any time (\textit{Liveness}), it can be considered as an almost strong random extractor. However, due to its dependency to history (although not very high), it was not able to pass several tests of Die Harder suite. Nevertheless, we advocate the evaluation steps used in this paper as a general evaluation method for PVS. The dataset we have used will be available at \cite{dataset}. We must mention that MRE-IPI should be computed in an IMD and it is using a far simpler and smaller algorithm where an IMD is very limited in every aspect of computation, yet, MRE-IPI was able to produce a quality as half as AES. We propose using MRE-IPI as a one-time pad in IMDs to receive the secure session key from the reader. This reduces the usage of MRE-IPI during a secure communication and thus it reduces the probability of breaking it.

\section{Acknowledgement}
This work has been funded by the UK EPSRC under grant EP/N023242/1 as part of the PETRAS Research Hub - Cyersecurity of the IoT.

\bibliographystyle{IEEEtran}
\bibliography{library}

\begin{thebibliography}{10}
\bibitem{Hamill:7nAp32ic}
J.~Hamill. {An inventor has fitted himself with a implant which replaces car
  keys and he says the future is HERE}.
\end{thebibliography}

\end{document}